\pdfoutput=1
\newcommand*{\ATLASLATEXPATH}{}
\documentclass[PAPER, atlasdraft=false, texlive=2016,cernpreprint, USenglish]{\ATLASLATEXPATH atlasdoc}
 
\usepackage[subfigure]{\ATLASLATEXPATH atlaspackage}
\usepackage[backref=false]{\ATLASLATEXPATH atlasbiblatex}
 
\usepackage[BSM,jetetmiss]{\ATLASLATEXPATH atlasphysics}
\usepackage{graphicx}
\usepackage{makecell}
\usepackage{ragged2e}
\usepackage{dcolumn}
\newcolumntype{d}[1]{D{.}{.}{#1}}

\graphicspath{{logos/}{figures/}}
 
\usepackage{ANA-SUSY-2018-06-PAPER-defs}

\AtlasTitle{Search for chargino-neutralino production with mass splittings near the electroweak scale in three-lepton final states
in $\sqrt{s} = 13$~\TeV\ $pp$ collisions with the ATLAS detector}
 
\AtlasAbstract{
A search for supersymmetry through the pair production of electroweakinos with mass splittings near the electroweak scale and decaying via on-shell $W$ and $Z$ bosons is presented for a three-lepton final state.
The analyzed proton--proton collision data taken at a center-of-mass energy of $\sqs = 13$~\TeV\ were collected between 2015 and 2018 by the ATLAS experiment at the Large Hadron Collider, corresponding to an integrated luminosity of 139~\ifb.
A search, emulating the recursive jigsaw reconstruction technique with easily reproducible laboratory-frame variables, is performed. The two excesses observed in the 2015--2016 data recursive jigsaw analysis in the low-mass three-lepton phase space are reproduced. Results with the full data set are in agreement with the Standard Model expectations.
They are interpreted to set exclusion limits at 95\% confidence level on simplified models of chargino--neutralino pair production for masses up to 345~\GeV.
}
 
\author{ATLAS Collaboration}
 
\AtlasRefCode{ATLAS-SUSY-2018-06}
 
\PreprintIdNumber{CERN-EP-2019-263}

\AtlasCoverSupportingNote{RJ mimic supporting document}{https://cds.cern.ch/record/2646712}
 
\AtlasJournal{\PRD}
\AtlasJournalRef{Phys. Rev. D 101 (2020) 072001}
\AtlasDOI{10.1103/PhysRevD.101.072001}

\AtlasCoverAnalysisTeam{Sara~Caroline~Alderweireldt, Jeffrey~Rogers~Dandoy, George~Dyckes, Joe~Kroll, Elliot~Lipeles, Joana~Machado~Miguens, Joseph~Reichert, Elodie~Deborah~Resseguie, Evelyn~Thomson, Iacopo~Vivarelli}
 
\AtlasCoverEdBoardMember{Volker~Buescher, Christina~Potter~(chair), Sara~Kristina~Strandberg}
 
\AtlasCoverEgroupEditors{atlas-SUSY-2018-06-editors@cern.ch}
 
\AtlasCoverEgroupEdBoard{atlas-SUSY-2018-06-editorial-board@cern.ch}

\hypersetup{pdftitle={ATLAS document},pdfauthor={The ATLAS Collaboration}}
 
\begin{document}
 
\maketitle
 
\section{Introduction}
\label{sec:intro}
Supersymmetry (SUSY)~\cite{Golfand:1971iw,Volkov:1973ix,Wess:1974tw,Wess:1974jb,Ferrara:1974pu,Salam:1974ig} is a space-time symmetry that extends the Standard Model (SM), predicting the existence of new partners for each SM particle.
The new particles have quantum numbers identical to those of their partners with the exception of spin, with SM fermions having bosonic partners and SM bosons having fermionic partners.
This extension presents solutions to deficiencies in the SM, addressing the hierarchy problem~\cite{Sakai:1981gr,Dimopoulos:1981yj,Ibanez:1981yh,Dimopoulos:1981zb} and providing a candidate for dark matter as the lightest supersymmetric particle (LSP), which will be stable if $R$ parity~\cite{Farrar:1978xj} is conserved~\cite{Goldberg:1983nd,Ellis:1983ew}.
 
The electroweakinos consist of two generations of charginos ($\chinopmi$, $i=1,2$) and four generations of neutralinos ($\ninoi$, $i=1,2,3,4$), where the indices are ordered by ascending mass,  with the LSP assumed to be the lightest neutralino, \ninoone. The electroweakinos are formed from the mixing of the SUSY partners of the Higgs field (known as Higgsinos) with the SUSY partners of the electroweak gauge fields, the bino for the U(1) gauge field and winos for the $W$ fields.
 
\begin{figure}[h]
\centering
\subfigure[]   {\label{fig:C1N2treeWZ} \includegraphics[width=0.35\textwidth]{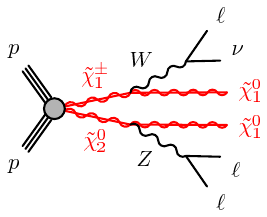}}
\subfigure[]{\label{fig:C1N2treeWZISR} \includegraphics[width=0.35\textwidth]{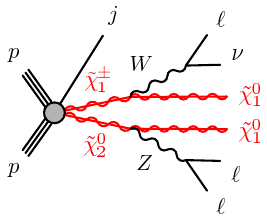}}
\caption{Diagrams of \chinoonepmninotwo production with subsequent
decays into two \ninoone and, via leptonically decaying $W$ and $Z$ bosons,  three leptons and a neutrino. Diagrams are shown both \subref{fig:C1N2treeWZ} without and \subref{fig:C1N2treeWZISR} with a jet from initial-state radiation.}
\label{fig:TreeDiagramsPhysScenarios}
\end{figure}

This paper presents a search for chargino-neutralino (\chinoonepmninotwo) pair production with
\chinoonepm--\ninoone and \ninotwo--\ninoone mass splittings near the electroweak scale.
The targeted decay chain is shown in Figure~\ref{fig:TreeDiagramsPhysScenarios}, with the chargino and neutralino decaying into the invisible LSP \ninoone and either a $W$ or $Z$ gauge boson, respectively. Simplified models~\cite{Alwall:2008ve,Alwall:2008ag,Alves:2011wf}, where the masses of the SUSY particles are the only free parameters, are used for interpretation. The \chinoonepm and \ninotwo are assumed to be purely wino and mass degenerate, and to decay with 100$\%$ branching ratio into $W$ and $Z$ bosons. The \ninoone LSP is assumed to be pure bino.
Both the $W$ and $Z$ bosons decay leptonically via SM branching ratios, leading to a final-state signature with three leptons and missing transverse momentum from two \ninoone and a neutrino.
The presence of initial-state radiation (ISR) may lead to jets in the final state and boost the \chinoonepmninotwo system, enhancing the signature of the missing transverse momentum.
The search targets a range of \chinoonepm/\ninotwo masses
between 100\footnote{A model-independent lower limit of 103.5 GeV at the 95$\%$ confidence level on the mass of promptly decaying charginos was set at the Large Electron-Positron Collider experiment~\cite{LEPlimits,Heister:2003zk,Abdallah:2003xe,Abbiendi:2003ji}.} and 450~$\GeV$ and mass splittings relative to the \ninoone LSP, $\Delta m = m({\chinoonepm/\ninotwo}) - m({\ninoone})$, larger than the $Z$ boson mass.
 
Previous searches for \chinoonepmninotwo production by the ATLAS~\cite{SUSY-2016-24,SUSY-2014-05,SUSY-2013-12} and CMS~\cite{CMS-SUS-16-039,CMS-SUS-17-004,CMS-SUS-13-006,CMS-SUS-12-006} collaborations using laboratory-frame, or conventional, observables in final states with two or three leptons, have found no significant excess of events in data over background expectations, yielding limits on the \chinoonepm and \ninotwo masses up to 580~\GeV\ and 570~\GeV, respectively.
A search by the ATLAS Collaboration using the recursive jigsaw reconstruction (RJR) technique~\cite{1607.08307,1705.10733} in $36.1~\ifb$ of data collected between 2015 and 2016~\cite{SUSY-2017-03} found excesses of three-lepton events in two regions. The excesses corresponded to a local significance of $2.1\sigma$ in the signal region targeting low-mass \chinoonepmninotwo production, \SRlow,  and to a local significance of $3.0\sigma$ in the signal region targeting \chinoonepmninotwo produced in association with ISR, \SRISR, with mass differences with respect to the LSP close to the $Z$ boson mass.
The signal regions of the RJR analysis targeting high-mass \chinoonepmninotwo production showed no substantial excess, with limits set on the \chinoonepm and \ninotwo masses of up to 600~\GeV\ for a massless \ninoone.
 
This new, independent analysis explores the intersection between the conventional and RJR approaches to better understand the tension between the exclusion limits produced by the two analyses.
It emulates the variables used by the RJR technique with conventional laboratory-frame discriminating variables, providing a simple set of variables that are easily reproducible. The object and region definitions using these new emulated recursive jigsaw reconstruction (eRJR) variables are kept as close as possible to those in Ref.~\cite{SUSY-2017-03}. This technique reproduces the three-lepton excesses in the low-mass region and ISR regions in the laboratory frame
using the same $36.1~\ifb$ of $pp$ collision data.
Results of this conventional search are also presented using a larger data set, corresponding to $139~\ifb$ of $pp$ collision data collected between 2015 and 2018.
 
A brief overview of the ATLAS detector is presented in Section~\ref{sec:detector}, and a description of the data set and the simulation of the \chinoonepmninotwo signal process and SM background processes is given in Section~\ref{sec:datamc}.
The reconstruction of the event and of objects used in the search is described in Section~\ref{sec:eventreco}. The eRJR kinematic discriminating variables are introduced in Section~\ref{sec:rjmethod}.
The search strategy is presented in Section~\ref{sec:search}, followed by the background estimation and validation in Section~\ref{sec:background}, and the systematic uncertainty derivation in Section~\ref{sec:systematics}.
The results of the search are presented in Section~\ref{sec:result}, followed by the conclusion in Section~\ref{sec:conclusion}.
 
\section{ATLAS detector}
\label{sec:detector}
The ATLAS detector~\cite{PERF-2007-01} is a multipurpose particle detector with almost 4$\pi$ coverage in solid angle.\footnote{
ATLAS uses a right-handed coordinate system with its origin at the nominal interaction point (IP) in the center of the detector and the \(z\)-axis along the beam pipe.
The \(x\)-axis points from the IP to the center of the LHC ring, and the \(y\)-axis points upwards.
Cylindrical coordinates \((r,\phi)\) are used in the transverse plane, \(\phi\) being the azimuthal angle around the \(z\)-axis.
The pseudorapidity $\eta$ is defined in terms of the polar angle \(\theta\) as \(\eta = -\ln \tan(\theta/2)\) and the rapidity $y$ is defined as $y= (1/2) \ln[(E+p_z)/(E-p_z)]$,
where $E$ is the energy and $p_z$ the longitudinal momentum of the physics object.
Angular distance is measured in units of \(\Delta R \equiv \sqrt{(\Delta\eta)^{2} + (\Delta\phi)^{2}}\), defined using $\eta$ unless otherwise specified.
}
It consists of an inner tracking system covering the pseudorapidity region $|\eta|<2.5$, sampling electromagnetic and hadronic calorimeters covering $|\eta|<4.9$, and a muon spectrometer covering $|\eta|<2.7$.
The inner detector (ID) reconstructs charged-particle tracks using silicon pixel and microstrip detectors and a straw-tube transition radiation tracker.
An additional innermost layer of the silicon pixel tracker, the insertable B-layer~\cite{ATLAS-TDR-19,PIX-2018-001}, was installed before 2015 at an average radial distance of 3.3~cm from the beam line to improve track reconstruction and flavor identification of quark-initiated jets.
The ID is surrounded by a thin, superconducting solenoid providing an axial magnetic field of 2~T, allowing the measurement of charged-particle momenta.
Beyond the ID is a high-granularity lead/liquid-argon (LAr) electromagnetic sampling calorimeter covering $|\eta|<3.2$ and a steel/scintillator-tile hadronic sampling calorimeter covering $|\eta|<1.7$.
The forward regions in $|\eta|$ are also covered by the copper/LAr hadronic end cap calorimeter ($1.7 < |\eta| < 3.2$) and by copper or tungsten/LAr forward calorimeters ($3.1 < |\eta| < 4.9$).
 
The muon spectrometer (MS) surrounds the calorimeters and measures muon tracks within a system of three superconducting air-core toroidal magnets with eight coils each.
The MS consists of three layers of precision tracking and triggering chambers.
 
The ATLAS trigger system consists of a hardware-based first-level (L1) trigger followed by a software-based high-level trigger (HLT)~\cite{TRIG-2016-01}.
The L1 trigger is designed to use a subset of detector information to accept events at an average rate of 100~kHz, while the HLT is designed to reduce the rate to an average of 1~kHz.
Candidate electrons are identified by the L1 trigger within the range $|\eta|<2.5$ as compact electromagnetic energy deposits in the electromagnetic calorimeter, and by the HLT using additional fast track reconstruction~\cite{TRIG-2018-05}.
Candidate muons are identified by the L1 trigger through a coincidence of MS trigger chamber layers and further selected by the HLT using fast reconstruction algorithms with input from the ID and MS.
 
 
\section{Data and Monte Carlo simulation}
\label{sec:datamc}
The data used for this search were collected between 2015 and 2018 by the ATLAS experiment and correspond to an integrated luminosity of $139~\ifb$.
The LHC collided protons at bunch-crossing intervals of 25~ns, with the average number of interactions per crossing measured in the data set to be $\avgmu=34$.
 
Monte Carlo (MC) simulation is used to model the expected contributions of various SM processes as well as possible SUSY signals.
The MC simulation is also used to optimize the event selection criteria and estimate the systematic uncertainties of the event yield measurement.
A full description of the MC simulation samples used is given below and summarized in Table~\ref{tab:MC}.
For most SM backgrounds, the expected contributions are taken from MC simulation, either directly or after normalization to data in dedicated control regions.
For \Zjets processes a data-driven method is used to predict the expected yield as described in Section~\ref{sec:background}, with MC simulation used in developing the method and estimating uncertainties.
 
Diboson, triboson, and  \Zjets samples~\cite{ATL-PHYS-PUB-2017-005,ATL-PHYS-PUB-2017-006} were simulated with the \SHERPAV{2.2}\cite{Bothmann:2019yzt} generator.
Diboson samples include fully leptonic and semileptonic decays as well as loop-induced and electroweak $VVjj$ production, where $V$ refers to a $Z$ or $W$ vector boson and $j$ indicates a jet.
The $WH$ and $ZH$ processes, with the Higgs boson ($H$) decaying into two $W$ or two $Z$ bosons, are included in the diboson and triboson samples.
The fully leptonic, the loop-induced, and the electroweak $VVjj$ diboson processes were simulated with \SHERPAV{2.2.2}, while the triboson, $Z$+jets, and semileptonically decaying diboson samples were simulated with \SHERPAV{2.2.1}.
 
In the \SHERPA samples the additional hard parton emissions~\cite{Gleisberg:2008fv} were matched to parton showers based on Catani-Seymour dipole factorization~\cite{Schumann:2007mg}.
The \nnpdfnnlo~\cite{Ball:2014uwa} set of parton distribution functions (PDFs) and a dedicated set of tuned parton-shower parameters (tune) developed by the \SHERPA authors were used~\cite{Schumann:2007mg}.
The matching of the matrix element to the parton shower~\cite{Hoeche:2011fd,Hoeche:2012yf,Catani:2001cc,Hoeche:2009rj} was employed for the various jet multiplicities, which were then merged into an inclusive sample using an improved CKKW matching procedure~\cite{Catani:2001cc} that is extended to next-to-leading-order (NLO) accuracy using the MEPS@NLO prescription~\cite{Hoeche:2012yf}.
The virtual QCD correction for matrix elements at NLO accuracy was provided by the \openloops\ library~\cite{Cascioli:2011va,Denner:2016kdg}.
The \Zjets\ (diboson) simulations were calculated for up to two (one) additional partons at NLO and up to four (three) additional partons at LO, while the triboson simulations were calculated for up to one additional parton at LO.
The cross sections predicted by these event generators were used for all samples except for the $Z$+jets processes, which were normalized to a next-to-next-to-leading-order (NNLO) cross section prediction~\cite{Anastasiou:2003ds}.
 
The production of \ttbar~\cite{Frixione:2007nw}, $\ttbar H$~\cite{Hartanto:2015uka}, and single-top $tW$~\cite{Re:2010bp}, \schan~\cite{Alioli:2009je}, and \tchan~\cite{Frederix:2012dh} events was modeled using the \POWHEGBOX~\cite{Nason:2004rx,Frixione:2007vw,Alioli:2010xd}~v2 generator at NLO with the \nnpdfnnlo PDF set.
The events were interfaced with \PYTHIAV{8.230}~\cite{Sjostrand:2014zea} using the A14 tune~\cite{ATL-PHYS-PUB-2014-021} and the \nnpdftwo PDF set~\cite{Ball:2012cx}.
The \hdamp\ parameter\footnote{The \hdamp\ parameter controls the transverse momentum \pt\ of the first additional emission
beyond the leading-order Feynman diagram in the parton shower and therefore regulates the  high-\pt\ emission against which the \ttbar\ system recoils.}
was set to 1.5 times the top-quark mass~\cite{ATL-PHYS-PUB-2016-020}.
The \ttbar\ inclusive production cross section was corrected to the theory prediction at NNLO in QCD including the resummation of next-to-next-to-leading-logarithmic (NNLL) soft-gluon terms calculated using \textsc{Top++2.0}~\cite{Czakon:2011xx}.
The $tW$ inclusive cross section was corrected to the theory prediction calculated at NLO in QCD with NNLL soft-gluon corrections~\cite{Aliev:2010zk,Kant:2014oha}.
Samples were generated in the five-flavor scheme, setting all quark masses to zero except for the top quark, and a diagram removal strategy~\cite{Frixione:2008yi} was employed in the $tW$ sample to handle the interference with \ttbar\ production~\cite{ATL-PHYS-PUB-2016-020}.
 
\begin{table}[t]
\footnotesize
\centering{
\caption{
Monte Carlo simulation details by physics process.  Listed are the generators used for matrix element calculation and for parton showering, the underlying-event parameter tunes, the PDF sets, and the order in \alphas of cross section calculations used for the yield normalization.
``Other top'' includes \tZ, \tWZ, $\ttbar WZ$, $\ttbar WW$, three-top, four-top, and rare top decay \ttbar ($t\rightarrow Wb \ell\ell)$ processes.
}
\label{tab:MC}
\begin{tabular}{lcccc}
\toprule
Process & Event     & PS and        & UE tune & Cross section \\
& generator & hadronization &         &               \\
\hline
\hline
 
\chinoonepmninotwo  & \MADGRAPHV{2.6}   & \PYTHIAV{8}   & A14     & NLO+NLL \\
Diboson             & \SHERPAV{2.2}     & \SHERPAV{2.2} & Default & NLO  \\
Triboson            & \SHERPAV{2.2}     & \SHERPAV{2.2} & Default & LO \\
$Z$+jets            & \SHERPAV{2.2}     & \SHERPAV{2.2} & Default & NNLO \\
\ttbar              & \POWHEGBOX~v2     & \PYTHIAV{8}   & A14     & NNLO+NNLL \\
Single~top          & \POWHEGBOX~v2     & \PYTHIAV{8}   & A14     & NLO+NNLL \\
$\ttbar H$          & \POWHEGBOX~v2     & \PYTHIAV{8}   & A14     & NLO \\
$\ttbar V$          & \MGMCatNLOV{2}  & \PYTHIAV{8}   & A14       & NLO \\
Other top           & \MGMCatNLOV{2}  & \PYTHIAV{8}   & A14       & LO \\
 
\bottomrule
\end{tabular}
}
\end{table}
 
The production of other top-quark processes was modeled using the \mgamc~v2~\cite{Alwall:2014hca} generator with the \nnpdfnnlo PDF set for the calculation of the matrix elements, which were interfaced with \PYTHIAV{8} using the A14 tune and the \nnpdftwo PDF set.
Generator versions \mgamc~v2.2.2 and \PYTHIAV{8.186} were used for $\ttbar WW$, 3-top, and 4-top processes, while
\mgamc~v2.3.3 and \PYTHIAV{8.212} were used for \tZ, \tWZ, $\ttbar V$, and $\ttbar WZ$ processes, as well as for \ttbar events that include the rare top decay $t\rightarrow Wb \ell\ell$.
All these top-quark processes were generated at LO except for $\ttbar V$, which was generated at NLO.
 
The SUSY \chinoonepmninotwo signal events were produced with up to two additional partons at LO using \mgamc~v2.6.1 with the \nnpdftwo PDF set, and were interfaced with \PYTHIAV{8.230}
using the A14 tune and \nnpdftwo PDF set.
The scale parameter for jet--parton CKKW-L matching was set to a quarter of the \chinoonepm/\ninotwo mass.
Signal cross sections were calculated at NLO in \alphas, adding the resummation of soft-gluon emission at next-to-leading-logarithm accuracy (NLL)~\cite{Beenakker:1999xh,Debove:2010kf,Fuks:2012qx,Fuks:2013vua,Fiaschi:2018hgm}. The nominal cross section and the uncertainty are taken from an envelope of cross section predictions using different PDF sets and factorization and renormalization scales~\cite{Borschensky:2014cia}. The inclusive cross section for \chinoonepmninotwo production, when each has a mass of 200~\GeV, is $1.8 \pm 0.1$~pb.
 
The decays of $c$ and $b$ hadrons in samples generated with \MGMCatNLO or \POWHEGBOX  were modeled with \evtgen~1.2.0~\cite{Lange:2001uf}.
Events from all generators were propagated through a full simulation of the ATLAS detector~\cite{SOFT-2010-01}
using \textsc{Geant4}~\cite{Agostinelli:2002hh}, which describes the interactions of particles with the detector.
A parameterized simulation of the ATLAS calorimeter called Atlfast-II~\cite{SOFT-2010-01}
was used for faster detector simulation of signal samples and is found to agree well with the full simulation.
The effect of multiple interactions in the same and neighboring bunch crossings (pileup) was modeled by
overlaying each hard-scattering event with simulated minimum-bias events generated with \PYTHIAV{8.210}
using the A3 tune~\cite{ATL-PHYS-PUB-2016-017} and \nnpdftwo PDF set.
 
\section{Event reconstruction}
\label{sec:eventreco}
Analysis events were recorded during stable beam conditions and must pass detector and data quality requirements.
Each event is required to have a primary vertex that is associated with a minimum of two tracks of transverse momentum $\pt > 500~\MeV$,
where the primary vertex is defined as the reconstructed vertex with the largest $\Sigma\pt^2$ of associated tracks~\cite{ATL-PHYS-PUB-2015-026}.
 
Two identification levels are defined for leptons and jets, referred to as ``baseline'' and ``signal,'' with signal objects being a subset of baseline.
The baseline leptons are required to satisfy looser identification and isolation criteria, providing a higher selection efficiency for leptons and jets for use in
calculating missing transverse momentum (\pTmiss), resolving ambiguities between overlapping physics objects, and calculating the data-driven estimate of the background arising from fake or nonprompt leptons.
 
Electron candidates are reconstructed using energy clusters in the electromagnetic calorimeter which are matched to an ID track,
and they are calibrated $\insitu$ using $Z\rightarrow ee$ decays~\cite{EGAM-2018-01}.
Baseline electrons must have $\pt > 10~\GeV$ and fall within the ID acceptance, $|\eta| < 2.47$.
The electrons must also satisfy the ``loose likelihood'' quality criteria~\cite{EGAM-2018-01}.
The trajectory of baseline electrons must be consistent with the primary vertex to suppress electrons originating from pileup.
Therefore, the tracks associated with baseline electrons must have a longitudinal impact parameter relative to the primary vertex ($z_0$) such that $|z_0 \sin\theta|< 0.5~\text{mm}$.
Signal electrons are required to satisfy the tighter ``medium'' identification criteria and must be well isolated from additional activity, passing a \pt-dependent ``tight'' isolation requirement that imposes fixed requirements on the values of the isolation variables.
The isolation is measured within a cone of size $\DeltaR=0.2$ around the electron, and the amount of nonassociated calorimeter transverse energy and scalar sum of track \pt must both be below 6\% of the electron \pt.
Tracks are only considered by the isolation criteria if they are consistent with the primary vertex.
For track isolation, the cone size decreases linearly with \pt above $50~\GeV$ as the electron's shower becomes more collimated.
The track associated with each signal electron must also pass a requirement on the transverse-plane distance of closest approach to the beam line ($d_0$) such that $|d_0/\sigma_{d_0}|< 5$, where $\sigma_{d_0}$ is the uncertainty in the value of $d_0$.
 
Muon candidates are reconstructed from either ID tracks matched to track segments in the MS or from tracks formed from a combined fit in the ID and MS~\cite{PERF-2015-10},
and they are calibrated $\insitu$ using $Z\rightarrow \mu\mu$ and $\Jpsi\rightarrow \mu\mu$ decays~\cite{PERF-2015-10}.
Baseline muons must have $\pt > 10~\GeV$, have $|\eta| < 2.4$ and pass an impact parameter cut of $|z_0 \sin\theta|< 0.5~\text{mm}$.
Signal muons must meet the ``medium'' identification criteria~\cite{PERF-2015-10} and the ``tight'' isolation criteria,
defined similarly to those for electrons but rejecting candidates with nonassociated calorimeter energy above 15\% and nonassociated track \pt above 4\% of the muon \pt.
The size of the track-isolation cone is $\DeltaR=0.3$ for muons with $\pt \leq 33~\GeV$ and decreases linearly to $\DeltaR=0.2$ at $\pt=50~\GeV$, improving the selection efficiency for higher-\pt muons.
The track associated with each signal muon must pass an impact parameter requirement of $|d_0/\sigma_{d_0}|< 3$.
 
Jet candidates are reconstructed from three-dimensional topological energy clusters~\cite{PERF-2014-07} using the \antikt algorithm~\cite{Fastjet,Cacciari:2008gp} with radius parameter $R=0.4$.
The jet energy scale (JES) and resolution (JER) are first calibrated to particle level using MC simulation and then $\insitu$ through $Z+$jet, $\gamma+$jet, and multijet measurements~\cite{PERF-2016-04}.
Baseline jets are required to have $\pt>20~\GeV$ and fall within the full calorimeter acceptance of $|\eta|<4.5$.
To suppress jets originating from pileup, jets are required to pass the ``medium'' working point of the track-based jet vertex tagger~\cite{PERF-2014-03,ATLAS-CONF-2014-018} if the jet has $\pt<120~\GeV$ and falls within the ID acceptance of $|\eta|<2.5$.
Signal jets are required to have $|\eta|<2.4$ to ensure full application of the pileup suppression, and events are rejected if they contain a jet that fails to meet the ``loose'' quality criteria~\cite{ATLAS-CONF-2015-029}, reducing contamination from noise bursts and noncollision backgrounds.
 
The identification of jets containing $b$ hadrons, called $b$ jets,  is performed using a multivariate discriminant built with information from track impact parameters, the presence of displaced secondary vertices, and the reconstructed flight paths of $b$ and $c$ hadrons inside the jet~\cite{FTAG-2018-01}.
The identification criteria are tuned to an average identification efficiency of 77\% as obtained for $b$ jets in simulated $\ttbar$ events,
corresponding to rejection factors of 110, 4.9, and 15 for jets originating from light quarks and gluons, $c$ quarks, and $\tau$ leptons, respectively.
 
To avoid reconstructing a single detector signature as multiple leptons or jets, an overlap removal procedure is applied to baseline leptons and jets.
For overlap removal, $\Delta R$ is calculated using rapidity, rather than $\eta$, to ensure the distance measurement is Lorentz invariant for jets that may have non-negligible masses.
First, any electron that shares a track with a muon in the ID is removed, as the track is seen to be consistent with track segments in the MS.
Then, jets are removed if they are within $\Delta R = 0.2$ of a lepton, as they have likely formed from an electron shower or muon bremsstrahlung.
For the overlap with associated muons, the nearby jet is discarded only if it is associated with less than three tracks of $\pt \ge 500~\MeV$.
Finally, electrons and muons with $\pt \leq 50~\GeV$ that are close to a remaining jet are discarded to reject nonprompt or fake leptons originating from hadron decays.
Leptons with $\pt \leq 25~\GeV$ are discarded if their distance from a jet is $\Delta R < 0.4$; for larger lepton \pt values up to 50~\GeV\ the $\Delta R$ discard range decreases linearly to $\Delta R < 0.2$.
 
The missing transverse momentum \pTmiss, with magnitude \met, is calculated as the negative vector sum of the transverse momenta of the baseline leptons, jets, and the soft term,
the latter given by the sum of the transverse momenta of additional low-momentum objects in the event~\cite{PERF-2016-07}.
The soft term is reconstructed from particle tracks in the ID that are associated with the primary vertex but not with any reconstructed analysis objects.
 
Data events were collected with triggers requiring either two electrons, two muons or an electron plus a muon.
The triggers have lepton \pt thresholds in the range 8--22~\GeV, and higher \pt thresholds are applied offline to ensure that the trigger efficiencies are constant in the relevant phase space.
All MC simulation samples emulate the triggers and have corrections applied to account for small differences with data in lepton identification, reconstruction, isolation and triggering efficiencies, as well as in jet pileup rejection and flavor identification efficiencies.
 
\section{Kinematic discriminants}
\label{sec:rjmethod}
In most $R$-parity-conserving SUSY models, the LSP is an invisible particle that rarely, if ever, interacts with matter.
It is therefore not directly observed by the ATLAS detector, but manifests itself as missing transverse momentum in an event whose particle transverse momenta would otherwise balance.
The relative boost of quarks in the colliding protons makes it impossible to know the true vector of the missing momentum, allowing only the transverse component to be measured accurately.
For SUSY particles with multiple decay steps the loss of this information can make it difficult to match the decay products and correctly reconstruct the originally produced particles,
resulting in ambiguities in the reconstruction of the \chinoonepm and the \ninotwo.
 
The RJR technique~\cite{1607.08307,1705.10733} attempts to resolve these ambiguities by analyzing each event starting from the laboratory-frame particles and boosting back to the rest frames of the parent particles.
Reconstructed jets, muons, and electrons are used as inputs for the RJR algorithm, which determines which leptons originate from the chargino or neutralino decays, assuming a specific decay chain.
The ISR jets are selected by minimizing the invariant mass of the system formed by the potential ISR jets and the sparticle system (consisting of the leptons and the missing-momentum vector) in the center-of-mass frame.
The only unknowns are the masses and longitudinal momenta of the invisible objects (two neutralinos and a neutrino), and how each individually contributes to the total missing energy.
The RJR algorithm determines the smallest Lorentz-invariant function of the visible particles' four-momenta that results in non-negative mass parameters for the invisible particles~\cite{1705.10733}.
 
This search targets \chinoonepmninotwo signals using an ISR region requiring the presence of one or more jets and a low-mass region with a jet-veto.
The eRJR technique emulates the RJR variables by using minimal assumptions about the mass of the invisible system and calculates all kinematic variables in the laboratory frame.
 
The eRJR variables, with original RJR variable names from Ref.~\cite{SUSY-2017-03}, used to select the ISR regions are defined as follows:
\begin{itemize}
\item \met: \pTI, the $\pt$ of the invisible particles, is emulated as the magnitude of the missing transverse momentum.
\item \pTjets: \pTisr, the $\pt$ in the vector sum of the ISR jets' momenta. In the eRJR technique, the ISR system  includes all signal jets in the event.
\item \dPhiMetJets: \dPhiISRMET, the azimuthal angle between the ISR system and the invisible particles, is emulated using the missing transverse momentum, $\pTmiss$, and the vector sum of the signal jets' momenta.
\item \RMetJets: \RISR, the normalized projection of the invisible system onto the ISR system, representing a ratio of \pTmiss to total jet \pt, is emulated as $| \pTmiss \cdot \hat{\mathbf{p}}_\text{T}^\text{jets} | / \pTjets$, where $\hat{\mathbf{p}}_\text{T}^\text{jets}$ is the unit vector of the vector sum of the signal jets' transverse momenta.
\item \pTsoft: \pTcm, the transverse momentum in the center-of-mass frame, where the ISR system recoils against the system containing the leptons and the missing energy, is emulated as the $\pt$ in the vector sum of the four-momenta of the signal jets, leptons, and \pTmiss, and is highly correlated with the \met soft term, defined in Section~\ref{sec:eventreco}.
\end{itemize}
 
Similarly, the eRJR variables, with original RJR variable names from Ref.~\cite{SUSY-2017-03} in parentheses, used in the low-mass regions are defined as follows:
\begin{itemize}
\item \pTsoft: \pTpp, the transverse momentum in the rest frame of the pair-produced sparticles (PP), is emulated as the $\pt$ in the vector sum of the four-momenta of the signal leptons and \pTmiss, and is identical to that of the ISR region except for the jet veto applied to the low-mass region.
\item \Ht: \HTppthreeone, the scalar sum of the $\pt$ of the signal leptons and the invisible system (neutrino and LSPs) in the PP frame, is emulated as the scalar sum of the $\pt$ of the signal leptons and \met.
\item \Hboost: \Hppthreeone, the scalar sum of the magnitude of the momenta of the signal leptons and the invisible system (neutrino and LSPs) in the PP frame, is emulated as the scalar sum of the momentum of the signal leptons and the missing-momentum vector (which includes longitudinal and transverse components), $|\pmiss|$,  after applying a boost.
\end{itemize}
 
To calculate \Hboost, the longitudinal component of the missing-momentum vector, $\plongmiss$,  and the boost need to be determined. The \plongmiss variable is calculated as~\cite{1705.10733}
\begin{equation*}
\plongmiss = p_{V,||} \frac{|\pTmiss|}{\sqrt{(\mathbf{p}_{V,\text{T}})^2 + m_V^2}},
\end{equation*}
where $p_{V,||}$ is the $z$ component of the vector sum of the four-momenta of the three signal leptons,  $p_{V,\text{T}}$ is the magnitude of the transverse momentum in the vector sum of the four-momenta of the three leptons, and $m_V$ is the mass of the three-lepton system. The invariant mass of the system of invisible particles is assumed to be zero and does not appear in the equation.
The boost of the system can then be calculated as:
\begin{equation*}
\mathbf{\pmb{\beta}} = \frac{\mathbf{p}}{E} = \frac{ \mathbf{p}^{V}+\pmiss } { E^{V} + |\pmiss|},
\end{equation*}
where $\mathbf{p}^{V}$ is the vector sum of the three-momenta of the three leptons, calculated in the laboratory frame. This boost is applied to the three leptons and the $\pmiss$. These new objects are used in the calculation of \Hboost.
 
The eRJR technique was validated against the published RJR result~\cite{SUSY-2017-03} and was able to reproduce an excess similar to that seen in the RJR analysis with the data set collected in 2015 and 2016. In \SRlow, exactly the same data events were selected using the emulated variables as when using the RJR variables with similar background expectation. In \SRISR, because all signal jets are considered part of the ISR system in the eRJR method, additional data events were selected alongside a proportional increase in the expected number of background events, with the significance of the excess in agreement with the RJR search. The signal significance in both the low-mass and ISR regions is comparable for both techniques.
A strong correlation is found between eRJR and RJR variables in loosened signal regions, with only a slight decorrelation seen in \pTsoft due to the differing jet selection, leading to the additional \SRISR events.
The eRJR technique provides a simple set of conventional variables in the laboratory frame that can easily be reproduced.
 
\section{Search strategy}
\label{sec:search}
This search is performed in signal regions (SRs) designed to select the targeted \chinoonepmninotwo signal events while accepting only a small but well-measured number of SM background events.
The SM background yields in the SRs are estimated using dedicated control regions (CRs) and confirmed in validation regions (VRs), as described in Section~\ref{sec:background}.
The full set of event selections is summarized in Table~\ref{tab:selections} and described below.
To target leptonically decaying $W$ and $Z$ bosons from the electroweakinos, events must have exactly three leptons which pass the baseline and signal requirements defined in Section~\ref{sec:eventreco}. The leptons must have at least one same-flavor opposite-charge (SFOS) pair ($e^+e^-$ or  $\mu^+\mu^-$) with an invariant mass \mll of the pair between 75~\GeV\ and 105~\GeV, consistent with the $Z$ boson mass. If there is more than one SFOS pair, the pair chosen is the one that has an invariant mass closest to the $Z$ boson mass.
 
\begin{table}[t]
\begin{center}
\caption{Selection criteria for the low-mass and ISR regions. The variables are defined in the text. In addition, events are required to have three signal leptons, and a $b$-jet veto is applied.
The invariant mass of the two leptons identified as coming from the $Z$ boson decay is between 75~\GeV\ and 105~\GeV, and the invariant mass of the three leptons is greater than 105~\GeV. }
\label{tab:selections}
\resizebox{\columnwidth}{!}{
\footnotesize
\setlength{\tabcolsep}{3.5pt}
\begin{tabular}{l | l l l l l l l l l l}
\hline
\multicolumn{10}{c}{Selection Criteria}\\
\hline
\hline
Low-mass Region & $p_\textrm{T}^{\ell_{1}}$ [\GeVNoSpace] & $p_\textrm{T}^{\ell_{2}}$ [\GeVNoSpace] & $p_\textrm{T}^{\ell_{3}}$ [\GeVNoSpace] & \njet & \mt [\GeVNoSpace]     & \met [\GeVNoSpace] & \Hboost [\GeVNoSpace] & \HTratio & \pTsoftratio\\
\hline
\CRlow          & $>60$                   & $>40$                   & $>30$           & $=0$        &  $\in(0,70)$    & $>40$       & $>250$         & $>0.75$  & $<0.2$      \\
\VRlow          & $>60$                   & $>40$                   & $>30$           & $=0$          &  $\in(70,100)$  & -           & $>250$         & $>0.75$  & $<0.2$      \\
\SRlow          & $>60$                   & $>40$                   & $>30$           & $=0$          &  $>100$         & -           & $>250$         & $>0.9$   & $<0.05$     \\
\hline
\hline
ISR Region           & $p_\textrm{T}^{\ell_{1}}$ [\GeVNoSpace] & $p_\textrm{T}^{\ell_{2}}$ [\GeVNoSpace] & $p_\textrm{T}^{\ell_{3}}$  [\GeVNoSpace] &  \njet & \mt [\GeVNoSpace] & \met [\GeVNoSpace] & \dPhiMetJets & \RMetJets        & \pTjets [\GeVNoSpace] & \pTsoft [\GeVNoSpace]  \\
\hline
\CRISR               &  $>25$                  & $>25$                   & $>20$                & $\geq 1$   & $<100$     & $>60$       & $>2.0$       & $\in(0.55,1.0)$  &  $>80$         & $<25$ \\
\VRISR               &  $>25$                  & $>25$                   & $>20$           & $\geq 1$         & $>60$      & $>60$       & $>2.0$       & $\in(0.55,1.0)$  &  $>80$         & $>25$ \\
\VRISRsmallptsoft    &  $>25$                  & $>25$                   & $>20$           & $\geq 1$         & $>60$      & $>60$       & $>2.0$       & $\in(0.55,1.0)$  &  $<80$         & $<25$ \\
\VRISRsmallrmetjets  &  $>25$                  & $>25$                   & $>20$         & $\geq 1$           & $>60$      & $>60$       & $>2.0$       & $\in(0.30,0.55)$ &  $>80$         & $<25$ \\
\SRISR               &  $>25$                  & $>25$                   & $>20$           & $\in[1,3]$        & $>100$     & $>80$       & $>2.0$       & $\in(0.55,1.0)$  &  $>100$        & $<25$ \\
\hline
\end{tabular}
}
\end{center}
\end{table}
 
The leading source of SM background is $WZ$ production, which for fully leptonic decays has three leptons and $\met$ from a neutrino in the final state.
To reduce the $WZ$ contribution, the transverse mass is calculated from the unpaired third lepton and the $\met$. It is defined as $\mt = \sqrt{2\pt\met~(1-\cos(\Delta\phi))}$, where $\Delta\phi$ is the azimuthal separation between the lepton and $\pTmiss$, and it will typically be at or below the $W$ boson mass in SM events where the \met is predominantly from the neutrino of the $W$ decay.
The \mt calculated in \chinoonepmninotwo events does not have such a constraint, and the SRs therefore require $\mt \ge 100~\GeV$ to reduce the SM $WZ$ background.
Additionally, signal events usually have larger values of $\met$ due to the undetected LSPs.
The backgrounds in which one or more leptons are fake or nonprompt are reduced by targeting the source of the additional leptons.
Events containing $b$ tagged jets are rejected to minimize contributions from the top backgrounds $t\bar{t}$ and $Wt$.
In the \Zjets background, a third signal lepton can arise from photon conversion, where the photon originates from the bremsstrahlung of a lepton off the $Z$ boson.
In this situation, all three signal leptons originate from the $Z$ boson, and this background can be reduced by requiring that the invariant mass \mlll of the three-lepton system be larger than 105~\GeV.
 
The signal regions are split into two different topologies: \SRlow, the low-mass region that requires a signal jet veto, and \SRISR, the ISR region that requires at least one central signal jet.
Both SRs are optimized for signals with small mass splittings, which can lead to events with lower-$\pt$ leptons or smaller $\met$ in the final state.
The inclusion of recoiling ISR boosts the invisible decay products in the same direction, enhancing the measured $\met$ and improving the discrimination between signal and the lower-$\met$ $WZ$ background.
 
The low-mass signal region requires the $\pt$ of the first, second, and third leptons (ordered in decreasing \pt) to be greater than 60~\GeV, 40~\GeV, and 30~\GeV, respectively, to minimize contributions from backgrounds with fake or nonprompt leptons.
Tight selection thresholds for \Hboost, \inlinepTsoftratio, and \inlineHTratio further reduce the $WZ$ contribution in the signal region.
The ISR region has a requirement of $\met \ge 80~\GeV$ to reduce the \Zjets background, which does not have a source of real $\met$.
The $\pt$ requirements on the three leptons can then be relaxed to be greater than 25~\GeV, 25~\GeV, and 20~\GeV, respectively, while ensuring that the dilepton triggers remain fully efficient.
To select the ISR topology in which the system of leptons and $\met$ is recoiling against the ISR jets, the azimuthal separation between the signal jets and \pTmiss, $\Delta\phi(\met,\text{jets})$, is required to be greater than 2.0.
The ratio of the $\pTmiss$ to the total transverse momenta of the jets is required to be $0.55 \le R(\met,\text{jets}) \le 1.0$ to ensure that most of the transverse momentum along the jet axis is carried by the invisible particles and not by the high-$\pt$ leptons from the $WZ$ background.
Requirements of $\pt^\text{soft}$ less than 25~\GeV\ and jet multiplicity $\njet$ less than four further reduce background contamination from $WZ$ events.

 
\section{Background estimation and validation}
\label{sec:background}
The backgrounds in this analysis can be classified into two groups:
irreducible backgrounds with at least three prompt leptons in the final state, and reducible backgrounds containing at least one fake or nonprompt lepton.
The dominant irreducible background is $WZ$ production, which is estimated by normalizing the yields in MC simulation to data in CRs using a log-likelihood fit described in Section~\ref{sec:result}.
Additional irreducible backgrounds include $ZZ$ and triboson production and processes that include a Higgs boson, three or more tops, and tops produced in association with a $W$ or $Z$ boson.
These backgrounds are estimated directly from MC simulation because of their small contribution.
The reducible backgrounds can be categorized into the top-quark-like \ttbar, $Wt$, and $WW$ processes, which are kinematically similar and mostly consist of nonprompt leptons from heavy-flavor hadron decays,
and the \Zjets process, which also accounts for the $Z$+$\gamma$ process, with fake or nonprompt leptons originating primarily from misidentified jets or photon conversions. The majority of the fake leptons from \Zjets are electrons from light flavor hadron decays, while few events are from photon conversions.
The reducible backgrounds are estimated separately in regions enriched in fake and nonprompt leptons, one region targeting top-quark-like processes and another targeting other fake/nonprompt sources, usually \Zjets processes, as described below.

The CRs for the $WZ$ background are designed to be kinematically similar but orthogonal to \SRlow and \SRISR.  They are enriched in $WZ$ events and the potential signal contamination is kept small (less than 10\% for all signal models).
To achieve this, an upper bound is placed on the $\mt$ of the CRs, targeting events that are likely to have a leptonically decaying $W$ boson and no other sources of real \met.
The low-mass CR (\CRlow) therefore requires $\mt<70$~\GeV\ while the ISR CR (\CRISR) has a slightly looser requirement of $\mt<100$~\GeV, benefiting from the boost of the \met system by the ISR.
The other kinematic selections are similar to those for the corresponding SRs but are also loosened to accept more $WZ$ events and reduce contamination from signal, as shown in Table~\ref{tab:selections}.
Figure~\ref{fig:CR} shows the background composition in the \CRlow and \CRISR regions, with good agreement seen between data and the background prediction after the fit.
 
\begin{figure}[h]
\centering
\subfigure[]{
\label{fig:CR:CRlow}
\includegraphics[width=0.48\textwidth]{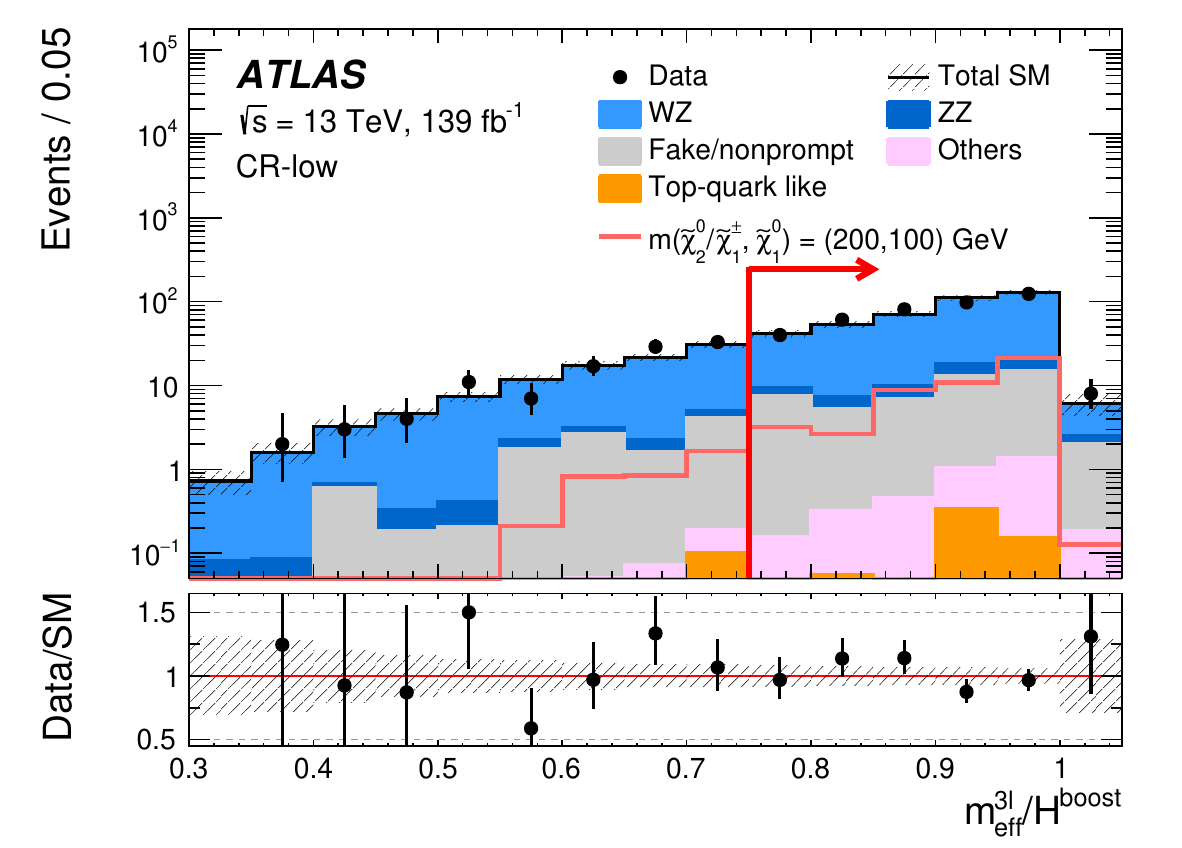}
}
\subfigure[]{
\label{fig:CR:CRISR}
\includegraphics[width=0.48\textwidth]{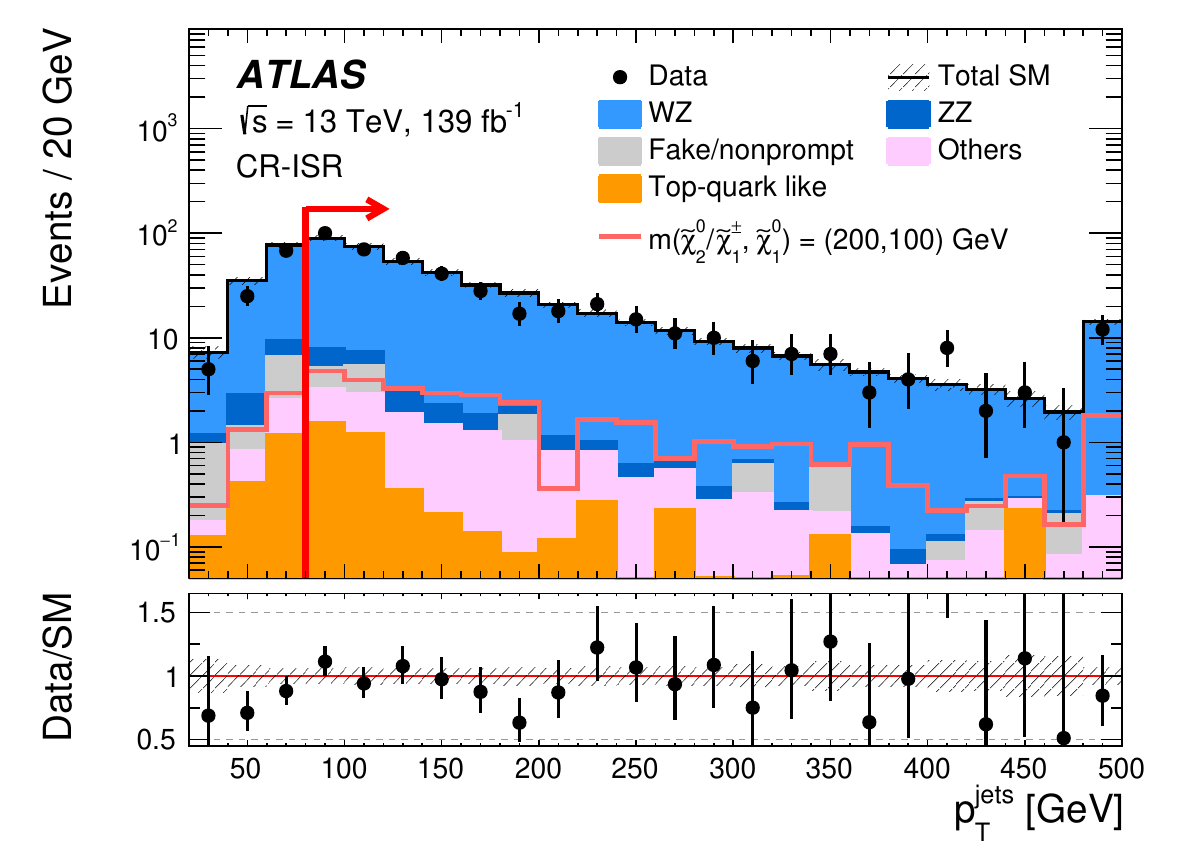}
}
 
\caption{Examples of kinematic distributions after the background-only fit, showing the data and the post-fit background in \subref{fig:CR:CRlow} \CRlow for $\Ht/\Hboost$ and
\subref{fig:CR:CRISR} \CRISR for \pTjets.
\captionsel{corresponding CR}
\captionstyle}
\label{fig:CR}
\end{figure}
 
The data-driven fake-factor method~\cite{STDM-2011-24,ATL-PHYS-PUB-2010-005} is used to estimate the fake/nonprompt-lepton background associated with the \Zjets process.
The fake-factor method uses two levels of lepton identification criteria. The regular identification (``reg-ID'') criteria correspond to the signal lepton criteria used in the analysis.
The reversed identification (``anti-ID'') criteria have one or more of the identification, isolation, or impact parameter criteria inverted relative to those of the signal leptons to obtain a selection enriched in fake leptons. A fake factor is then defined as the ratio of the yield of reg-ID leptons to the yield of anti-ID leptons in a given region of phase space.
The fake factors are measured in a region dominated by \Zjets events, requiring $\met < 40~\GeV$, $\mt<30~\GeV$, $|m_{\ell\ell}-m_{Z}| <15~\GeV$, and a $b$ jet veto.
The two leptons identified as the $Z$ boson decay products must pass the signal lepton requirements, while the unpaired lepton must satisfy either the reg-ID or anti-ID criteria.
Electron and muon fake factors are then measured separately and as a function of lepton $\pt$. They are validated in a statistically independent region with a similar selection but requiring $\met < 40$~\GeV\ and $30<\mt<50$~\GeV, so as to be closer to the signal region.
The derived fake factors are applied to events satisfying the same criteria as for the CRs, VRs, and SRs (defined in Table~\ref{tab:selections}), while additionally requiring that at least one of the signal leptons is replaced by an anti-ID lepton.
In both the derivation and application of the fake factors, the prompt lepton and top-quark-like backgrounds that have one or more anti-ID leptons are subtracted to avoid double counting.
 
The top-quark-like background contribution is estimated using MC simulation normalized to data in a top-quark-dominated CR. The region is constructed using different-flavor, opposite-charge ($e^\pm e^\pm \mu^\mp$ or $\mu^\pm \mu^\pm e^\mp$) trilepton events with lepton $\pt$ thresholds of 25, 25, and 20~\GeV\ as well as a $b$ jet veto. The normalization factors are applied to the same-flavor opposite-charge events in the top-quark-like MC simulation.
 
Four validation regions are designed in order to check that the background estimate agrees with data in regions kinematically closer to the SRs, typically targeting the extrapolation from CR to SR in a specific variable.
The full VR definitions are summarized in Table~\ref{tab:selections}. The VR definitions are also chosen to keep the contamination from signal below 10$\%$.
A low-mass validation region, \VRlow, is designed to test the extrapolation in $\mt$ between \CRlow and \SRlow, requiring $70<\mt<100$~\GeV.
Three ISR validation regions, \VRISR, \VRISRsmallptsoft, and \VRISRsmallrmetjets, invert different selections to validate the modeling in a varied phase space.
The total yields in the CRs and VRs are shown in Table~\ref{tab:results_CRVR_low} for the low-mass regions and in Table~\ref{tab:results_CRVR_ISR} for the ISR regions.
Figure~\ref{fig:VR} shows distributions in \VRlow, \VRISR, \VRISRsmallptsoft, and \VRISRsmallrmetjets for the full background prediction.
The background predictions and the observed data are generally in good agreement after the fit.
The data and background predictions in \VRlow and \VRISRsmallptsoft agree within $2\sigma$, and good agreement is seen in the shapes of relevant kinematic distributions.
 

\begin{table}
\begin{center}
\caption{
The observed and expected yields after the background-only fit in the low-mass CR and VR.
The normalization factors of the $WZ$ sample for the low-mass and ISR regions are different
and are treated separately in the combined fit.
\captioncategories
Combined statistical and systematic uncertainties are presented.
The individual uncertainties can be correlated and do not necessarily add in quadrature to equal the total background uncertainty.
}
\label{tab:results_CRVR_low}
\setlength{\tabcolsep}{0.0pc}
\gethalfcolwidth{colA}{\CRlow}
\gethalfcolwidth{colB}{\VRlow}
 
{\footnotesize
\renewcommand{\arraystretch}{1.1}
\begin{tabular*}{\textwidth}{
@{\extracolsep{\fill}}
l
R{\colA}@{\extracolsep{.16667em}}l@{\extracolsep{\fill}}
R{\colB}@{\extracolsep{.16667em}}l@{\extracolsep{\fill}}
}
\toprule
& \multicolumn{2}{l}{\CRlow}            & \multicolumn{2}{l}{\VRlow}              \\[-0.05cm]
\midrule
Observed events           &  \multicolumn{2}{c}{412}       &  \multicolumn{2}{c}{338} \\
\midrule\midrule
Fitted SM events          &  412  & $\pm$ 20               &  291  & $\pm$ 20 \\
\midrule
$WZ$                      &  343  & $\pm$ 27               &  262  & $\pm$ 22 \\
$ZZ$                      &  19.2 & $\pm$ 1.6              &  18.2 & $\pm$ 1.7 \\
Others                    &  3.0  & $\pm$ 1.5              &  1.6  & $\pm$ 0.8 \\
Top-quark like            &  0.5  & $\pm$ 0.4              &  0.02 & $_{-0.02}^{+0.25}$ \\
Fake/nonprompt leptons    &  46   & $\pm$ 17               &  10   & $\pm$ 5 \\
 
\bottomrule
\end{tabular*}
}
\end{center}
\end{table}

\begin{table}
\begin{center}
\caption{
The observed and expected yields after the background-only fit in the ISR CR and VRs.
The normalization factors of the $WZ$ sample for the low-mass and ISR regions are different
and are treated separately in the combined fit.
\captioncategories
Combined statistical and systematic uncertainties are presented.
The individual uncertainties can be correlated and do not necessarily add in quadrature to equal the total background uncertainty.
}
\label{tab:results_CRVR_ISR}
\setlength{\tabcolsep}{0.0pc}
\gethalfcolwidth{colA}{\CRISR}
\gethalfcolwidth{colB}{\VRISR}
\gethalfcolwidth{colC}{\VRISRsmallptsoft}
\gethalfcolwidth{colD}{\VRISRsmallrmetjets}
 
{\footnotesize
\renewcommand{\arraystretch}{1.1}
\begin{tabular*}{\textwidth}{
@{\extracolsep{\fill}}
l
R{\colA}@{\extracolsep{.16667em}}l@{\extracolsep{\fill}}
R{\colB}@{\extracolsep{.16667em}}l@{\extracolsep{\fill}}
R{\colC}@{\extracolsep{.16667em}}l@{\extracolsep{\fill}}
R{\colD}@{\extracolsep{.16667em}}l@{\extracolsep{\fill}}
}
\toprule
& \multicolumn{2}{c}{\CRISR}            &  \multicolumn{2}{c}{\VRISR}            &  \multicolumn{2}{c}{\VRISRsmallptsoft}            &  \multicolumn{2}{c}{\VRISRsmallrmetjets}              \\[-0.05cm]
\midrule
Observed events          & \multicolumn{2}{c}{442}              & \multicolumn{2}{c}{101}              & \multicolumn{2}{c}{72}              & \multicolumn{2}{c}{253}                    \\
\midrule\midrule
Fitted SM events          &  442 & $\pm$ 21               &  111 & $\pm$ 19               & 96  & $\pm$ 7                 &  256 & $\pm$ 13                 \\
\midrule
$WZ$                      &  415 & $\pm$ 22               &  98  & $\pm$ 17               & 89  & $\pm$ 7                 &  245 & $\pm$ 13                 \\
$ZZ$                      &  9.1 & $\pm$ 0.8              &  2.1 & $\pm$ 0.5              & 2.6 & $\pm$ 0.4               &  2.7 & $\pm$ 0.4                \\
Others                    &  12  & $\pm$ 6                &  6.9 & $\pm$ 3.5              & 1.7 & $\pm$ 0.9               &  6.2 & $\pm$ 3.2                \\
Top-quark like            &  4.7 & $\pm$ 1.6              &  2.7 & $\pm$ 1.1              & 1.5 & $\pm$ 1.2               &  2.0 & $\pm$ 1.0                \\
Fake/nonprompt leptons    &  1.5 & $_{-1.5}^{+2.3}$       &  0.9 & $_{-0.9}^{+1.6}$       & 1.3 & $_{-1.3}^{+1.6}$        & 0.01 & $_{-0.01}^{+0.05}$       \\
 
\bottomrule
\end{tabular*}
}
\end{center}
\end{table}

\begin{figure}[h]
\centering
\subfigure[]{
\label{fig:VR:VRlow}
\includegraphics[width=0.48\textwidth]{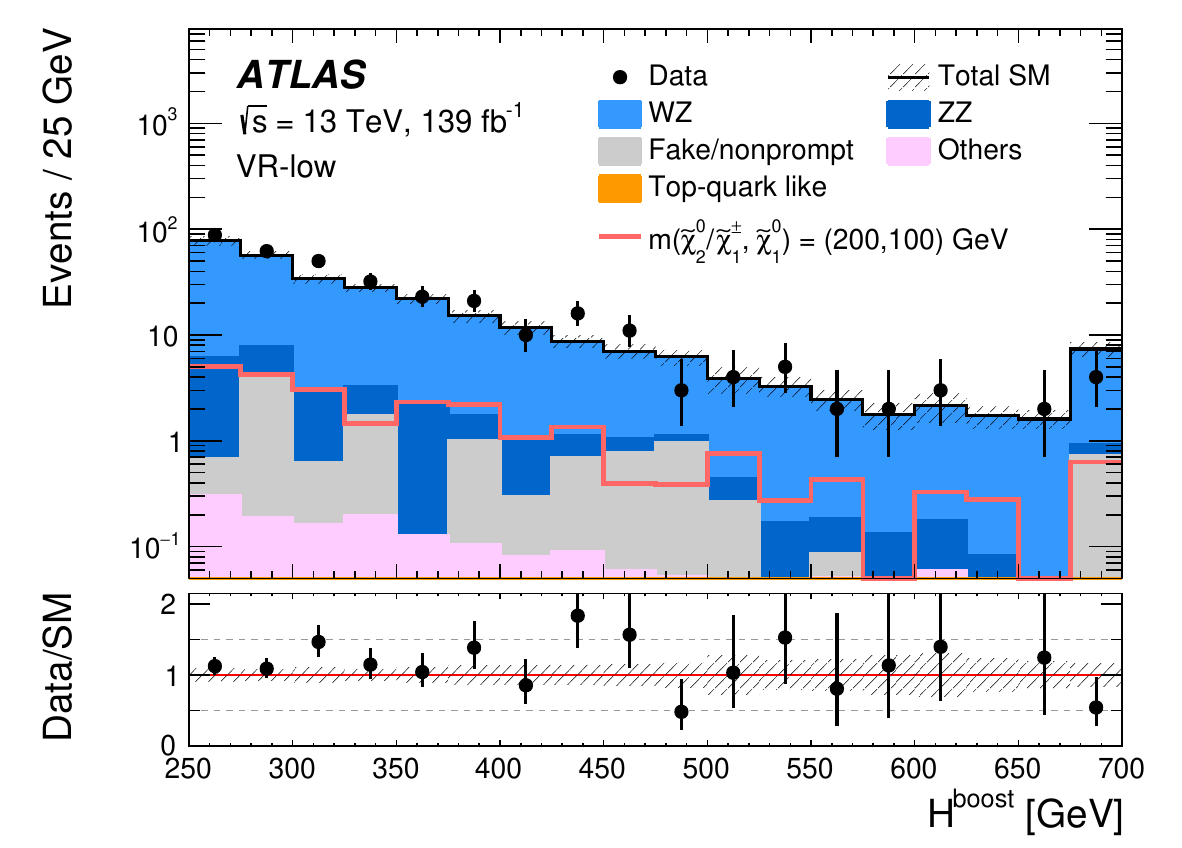}
}
\subfigure[]{
\label{fig:VR:VRISR}
\includegraphics[width=0.48\textwidth]{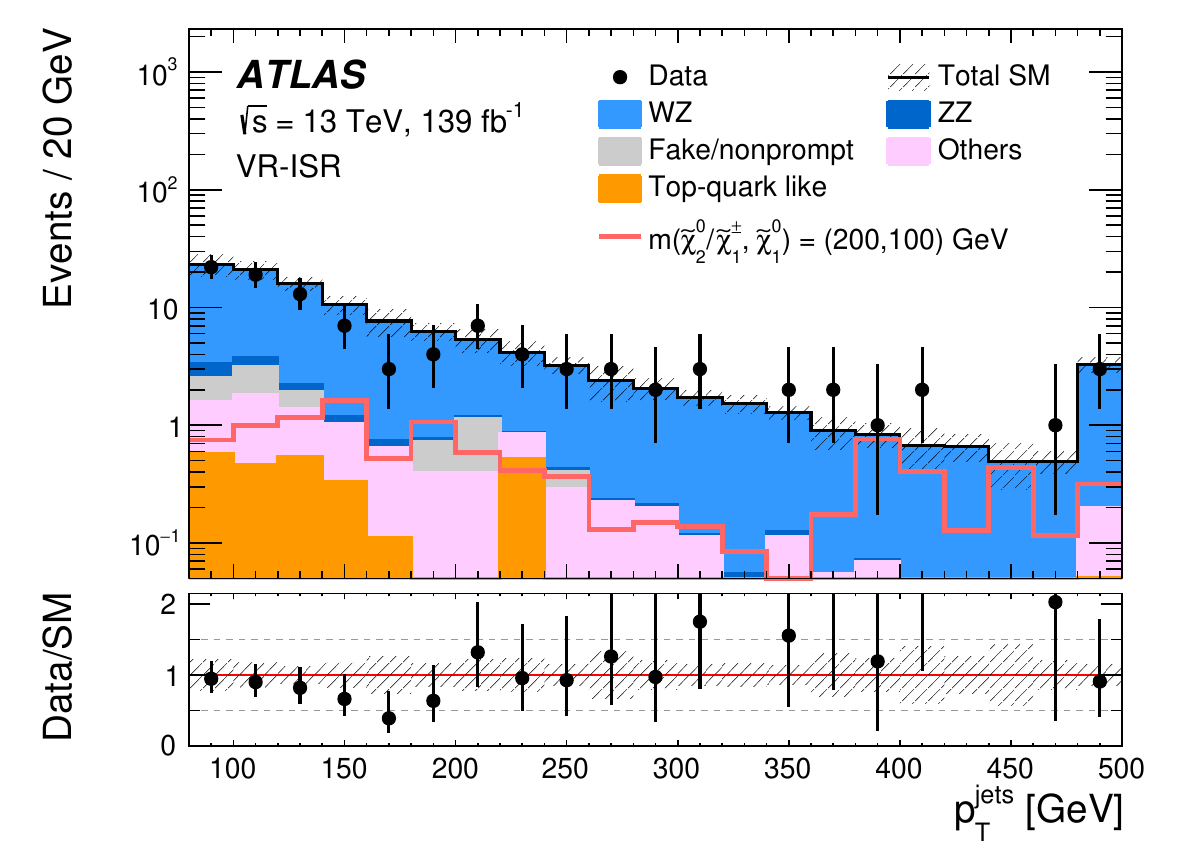}
}
\subfigure[]{
\label{fig:VR:VRISRsmallptsoft}
\includegraphics[width=0.48\textwidth]{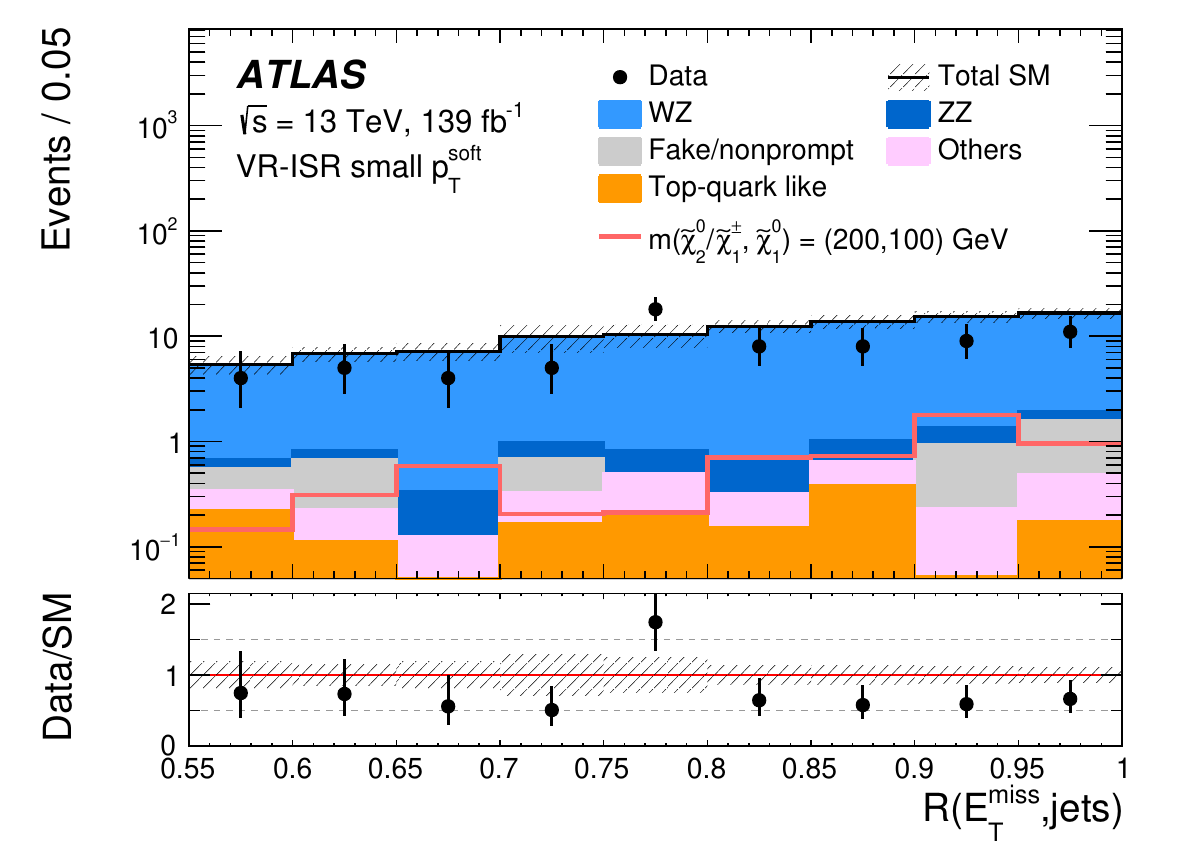}
}
\subfigure[]{
\label{fig:VR:VRISRsmallrmetjets}
\includegraphics[width=0.48\textwidth]{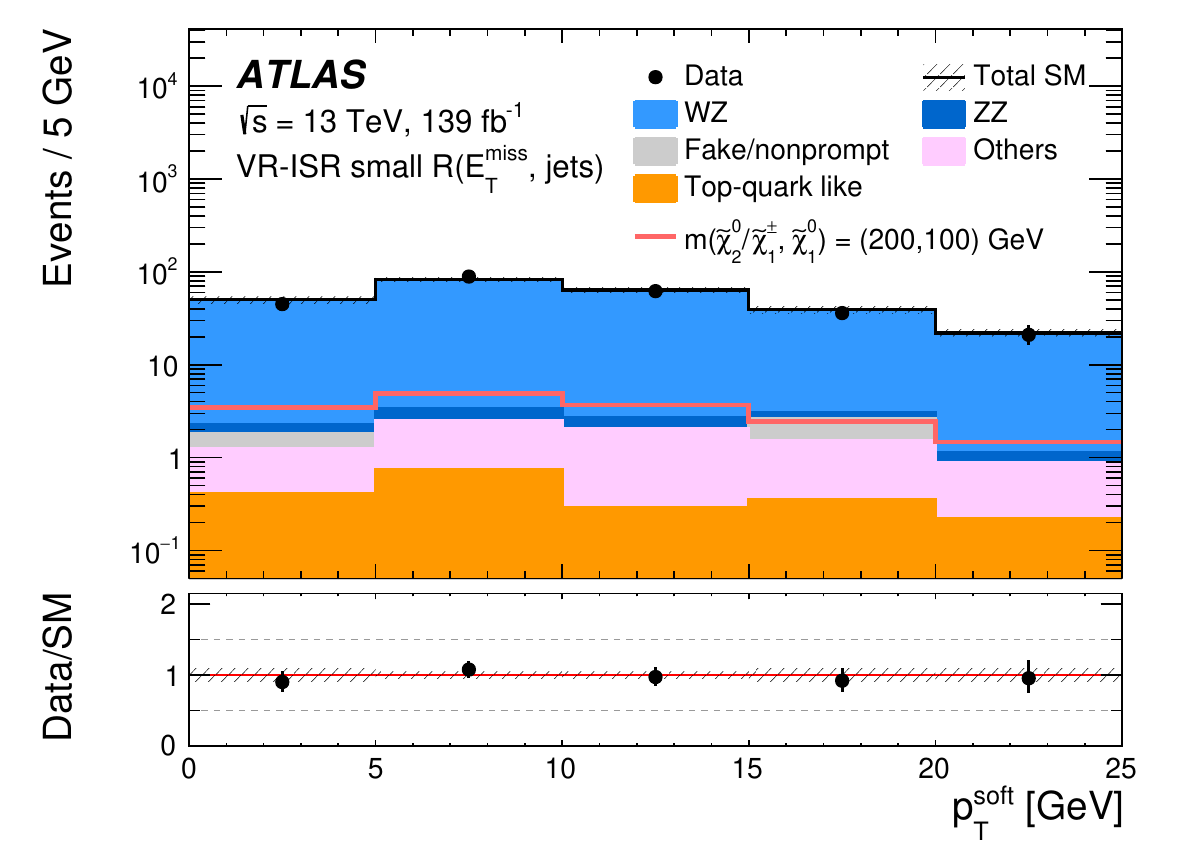}
}
 
\caption{Kinematic distributions showing the data and post-fit background in \subref{fig:VR:VRlow} \VRlow for \Hboost, \subref{fig:VR:VRISR} \VRISR for \pTjets,
\subref{fig:VR:VRISRsmallptsoft} \VRISRsmallptsoft for \RMetJets, and \subref{fig:VR:VRISRsmallrmetjets} \VRISRsmallrmetjets for \pTsoft.
\captionstyle}
\label{fig:VR}
\end{figure}
 
\section{Systematic uncertainties}
\label{sec:systematics}
Systematic uncertainties are derived for the signal and background predictions and include experimental uncertainties in detector measurements as well as theoretical uncertainties in the expected yields and MC simulation modeling.
 
 
\begin{table}
\begin{center}
\caption{Summary of the dominant experimental and theoretical uncertainties in the SM background prediction in the low-mass and ISR signal regions.
The individual uncertainties can be correlated and do not necessarily add in quadrature to equal the total post-fit background uncertainty.}
\label{tab:syst_summary}
\setlength{\tabcolsep}{0.0pc}
{
\footnotesize
\begin{tabular*}{\textwidth}{@{\extracolsep{\fill}}l *{2}{d{4.4}@{}}}
\toprule
\textbf{Uncertainty in signal regions}                                    & \multicolumn{1}{r}{\SRlow}            &  \multicolumn{1}{r}{\SRISR}            \\
\midrule
Jet energy scale and resolution         &  7.1 \%          & 6.1 \%       \\
$WZ$ normalization procedure        & 6.6 \%          & 4.5 \%       \\
\met         & 3.3 \%         & 2.1 \%      \\
MC statistics         & 2.9 \%          & 3.9 \%       \\
Anti-ID CR statistics         &2.7 \%        & 0.21 \%     \\
$WZ$ theory         & 1.9 \%        & 1.3 \%       \\
30\% uncertainty in minor backgrounds (``Others”)         &1.7 \%         & 3.1 \%       \\
Fake-factor estimation  & 1.2 \%          & < 0.01 \%    \\
Muon momentum scale and resolution         & 0.30 \%          &  0.04 \%       \\
Electron energy scale and resolution         &0.22 \%          & 0.34 \%       \\
Pileup         & 0.20 \%         &  0.94 \%       \\
Top-quark-like background estimation         &  < 0.01 \%          & 1.4 \%       \\
Flavor tagging         &  < 0.01 \%         & 0.47 \%       \\
\bottomrule
\end{tabular*}
}
\end{center}
\end{table}
 
Experimental uncertainties reflect the accuracy of the experimental measurements of jets, electrons, muons, and \met.
The JES and JER uncertainties~\cite{PERF-2016-04,PERF-2011-04} are derived as a function of jet \pt and $\eta$ and account for dependencies on the pileup conditions and on the flavor composition of jets.
The JES reflects the uncertainty in the average jet \pt measurement, varying from about 4\% for 20~\GeV\ jets to 1\% above 300~\GeV, while the JER reflects the uncertainty in the precision of the jet \pt measurement, varying from about 2\% to 0.4\% across the same \pt range.
Varying the JES and JER can alter the jet multiplicity of an event, affecting its inclusion in the \SRlow or \SRISR regions, as well as affecting the eRJR variables that depend on jet and \met kinematics.
Similar types of uncertainties account for the energy scales and resolutions for electrons~\cite{EGAM-2018-01} and muons~\cite{PERF-2015-10}.
Variations reflecting the per-object uncertainties are propagated through the \met calculation, with additional uncertainties for the scale and resolution of the soft term~\cite{PERF-2016-07}.
 
Additional experimental uncertainties account for differences between the data and MC simulation in the efficiency of the identification, reconstruction, isolation, and triggering of electrons~\cite{EGAM-2018-01} and muons~\cite{PERF-2015-10},
in the identification of pileup jets by the jet vertex tagger~\cite{PERF-2014-03},
and in the identification of $b$ jets by the flavor-tagging algorithm~\cite{FTAG-2018-01}.
An uncertainty on the pileup modeling is also considered and found to be small.
The uncertainty in the combined 2015--2018 integrated luminosity is 1.7\%~\cite{ATLAS-CONF-2019-021}, obtained using the LUCID-2 detector~\cite{LUCID2} for the primary luminosity measurements.
 
The theoretical uncertainties account for mismodeling in the MC simulation, particularly for the $WZ$ process. They include QCD scale uncertainties affecting the $WZ$ cross section, PDF uncertainties, and the uncertainty in \alphas.
The effects of QCD scale uncertainties are evaluated using seven-point variations of the factorization and renormalization scales in the matrix elements.
The scales are varied upwards and downwards by a factor of two, allowing for both independent and correlated variations of the two scales but prohibiting the anticorrelated variations.
The PDF uncertainties are evaluated by taking the envelope of the 100 variation replicas of the nominal PDF set and the central values of the \ctfourteennnlo~\cite{Dulat:2015mca} and \mmhtnnlo~\cite{Harland-Lang:2014zoa} PDF sets. The impact of $\pm 0.001$ shifts of $\alphas$ on the acceptance is also considered.
The QCD scale uncertainty is dominant and affects the prediction of the amount of additional radiation, and therefore the jet multiplicity, within an event.
The effect of the QCD scale uncertainty grows with the number of jets in an event, but the total uncertainty in the CR-to-SR transfer factor is reduced by similarities between the jet multiplicity distributions in the control and signal regions.
An additional uncertainty on the $WZ$ normalization procedure arises from the use of CRs to normalize the $WZ$ contributions in the SRs.
 
Uncertainties in the cross section are included for the signals and minor backgrounds whose yields are taken directly from MC simulation, with signal uncertainties varying with \chinoonepm/\ninotwo mass from 4.3\% at 100~\GeV\ to 11.5\% at 750~\GeV.
Uncertainties in the amount of initial- and final-state radiation are derived for each signal sample in the two signal regions by considering the ten eigenvariations of the A14~\cite{ATL-PHYS-PUB-2014-021} tune summed in quadrature, giving uncertainties of 15\%.
 
The systematic uncertainty of the data-driven fake/nonprompt (FNP) lepton estimate accounts for statistical uncertainties of the measured fake factors, assumptions made in the fake-factor method, and the closure of the method using MC simulation.
The number of MC simulation events with prompt leptons, primarily from $WZ$ events, that is subtracted in the fake-factor estimation is varied upwards and downwards by the $WZ$ cross section uncertainty of 5\%~\cite{ATLAS-CONF-2018-034}, leading to an uncertainty in the FNP lepton estimate of 6.7\% for electrons and 15.3\% for muons.
The nominal FNP lepton estimate is derived as a function of lepton \pt, and good agreement is generally seen for other kinematic variables in the FNP lepton control and validation regions.
A slight dependence on $|\eta|$ is seen, with typical deviations of 25\% for electrons and 21\% for muons taken as additional uncertainties.
The method closure uncertainty instead uses the \Zjets MC simulation to derive the FNP lepton estimate and compares the predictions in a loosened signal region with the known FNP lepton yield from MC simulation, accounting for potential differences in the FNP lepton composition between regions. Uncertainties of 12\% and 18\% are derived for electrons and muons, respectively.
The total impact of the fake-factor uncertainties is relatively small in both signal regions given the small contribution from backgrounds with FNP leptons.
 
The dominant uncertainties are summarized in Table~\ref{tab:syst_summary} for both the \SRlow and \SRISR regions.
The largest experimental uncertainties reflect the unknowns of the energy and \pt calibration of jets and the measurement of the soft term of the \met.
The largest theoretical source is the uncertainty in the QCD factorization and renormalization scales for the $WZ$ cross section.
The analysis also accounts for the statistical uncertainty of the MC simulation samples.


\section{Results}
\label{sec:result}
\begin{figure}[h!]
\centering
\includegraphics[width=0.95\textwidth]{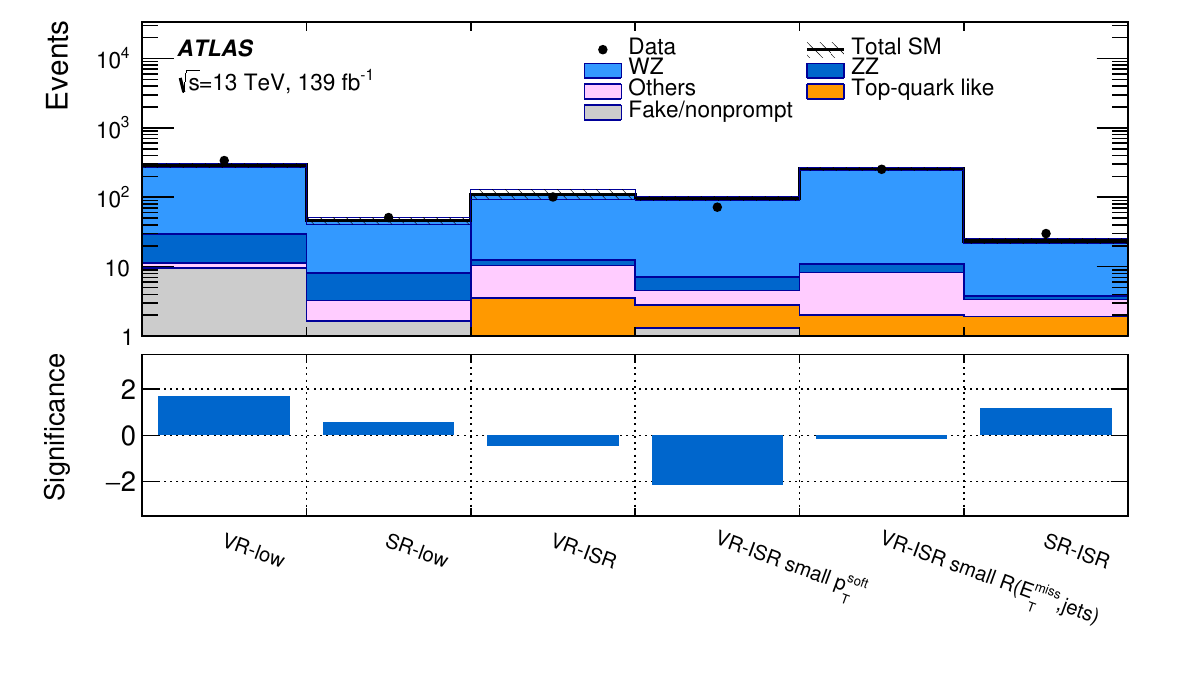}
\caption{ The observed data and expected SM background yields in the VRs and SRs.  The SM background prediction is derived with the background-only fit configuration, and the hatched band includes the experimental, theoretical, and statistical uncertainties.
\captioncategories
The bottom panel shows the significance~\cite{Cousins:2007bmb} of the differences between the observed and expected yields.}
\label{fig:region_yields}
\end{figure}


\begin{table}
\begin{center}
\caption{
The observed and expected yields after the background-only fit in the SRs.
The normalization factors of the $WZ$ sample for the low-mass and ISR regions are different
and are treated separately in the combined fit.
\captioncategories
Combined statistical and systematic uncertainties are presented.
The individual uncertainties can be correlated and do not necessarily add in quadrature to equal the total background uncertainty.
}
\label{tab:resultsSR}
\setlength{\tabcolsep}{0.0pc}
\gethalfcolwidth{colA}{\SRlow}
\gethalfcolwidth{colB}{\SRISR}
{\small
\renewcommand{\arraystretch}{1.1}
\begin{tabular*}{\textwidth}{
@{\extracolsep{\fill}}
l
R{\colA}@{\extracolsep{.16667em}}l@{\extracolsep{\fill}}
R{\colB}@{\extracolsep{.16667em}}l@{\extracolsep{\fill}}
}
\toprule
&  \multicolumn{2}{c}{\SRlow}    &  \multicolumn{2}{c}{\SRISR}              \\[-0.05cm]
\midrule
Observed events           &  \multicolumn{2}{c}{51}        &  \multicolumn{2}{c}{30}                  \\
\midrule\midrule
Fitted SM events          &  46   & $\pm$ 5                &  23.4 & $\pm$ 2.1             \\
\midrule
$WZ$                      &  38   & $\pm$ 5                &  19.7 & $\pm$ 2.0             \\
$ZZ$                      &  4.9  & $\pm$ 0.6              &  0.38 & $\pm$ 0.08            \\
Others                    &  1.6  & $\pm$ 0.8              &  1.5  & $\pm$ 0.8             \\
Top-quark like            &  0.03 & $_{-0.03}^{+0.18}$     &  1.9  & $\pm$ 0.8             \\
Fake/non-prompt leptons   &  1.6  & $\pm$ 1.3              &  0.01 & $_{-0.01}^{+0.05}$    \\
 
\bottomrule
\end{tabular*}
}
\end{center}
\end{table}
 
\begin{figure}[h]
\centering
\subfigure[]{
\includegraphics[width=0.48\textwidth]{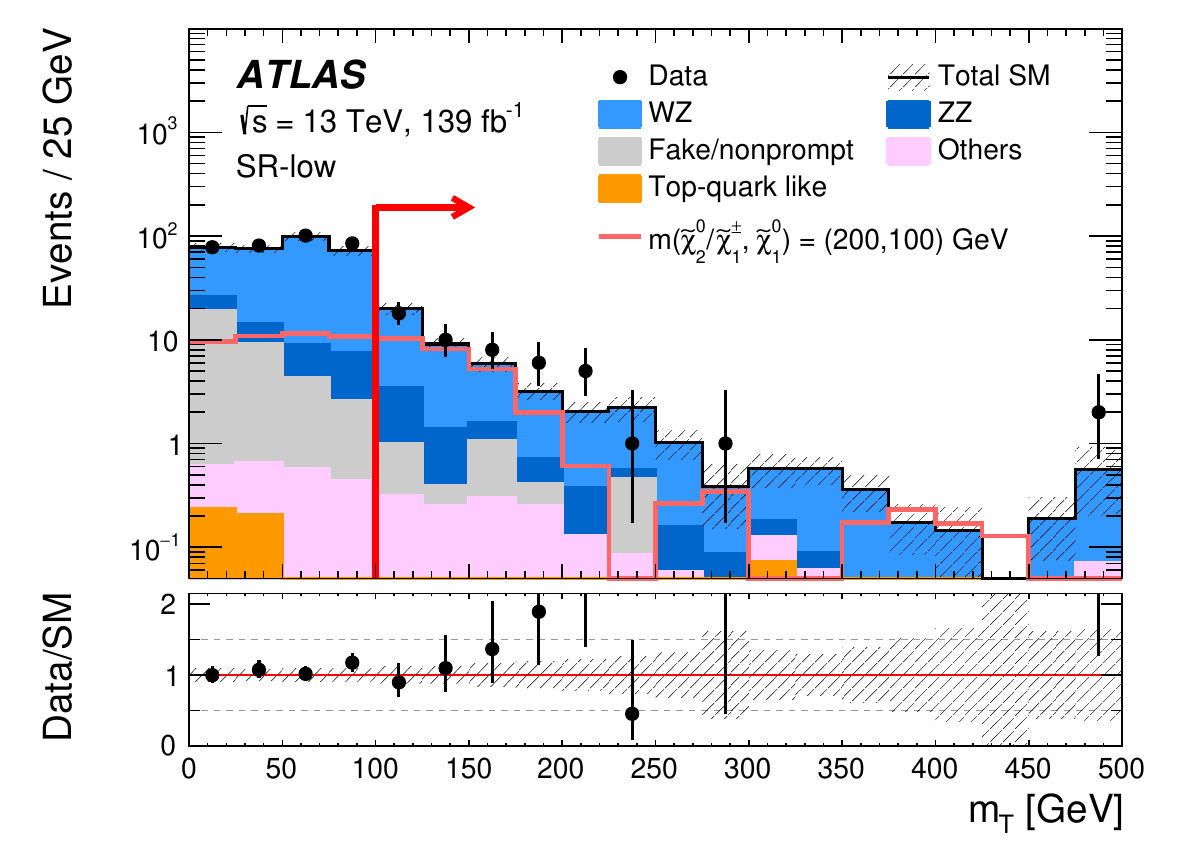}
\label{fig:SRLow:mt}
}
\subfigure[]{
\includegraphics[width=0.48\textwidth]{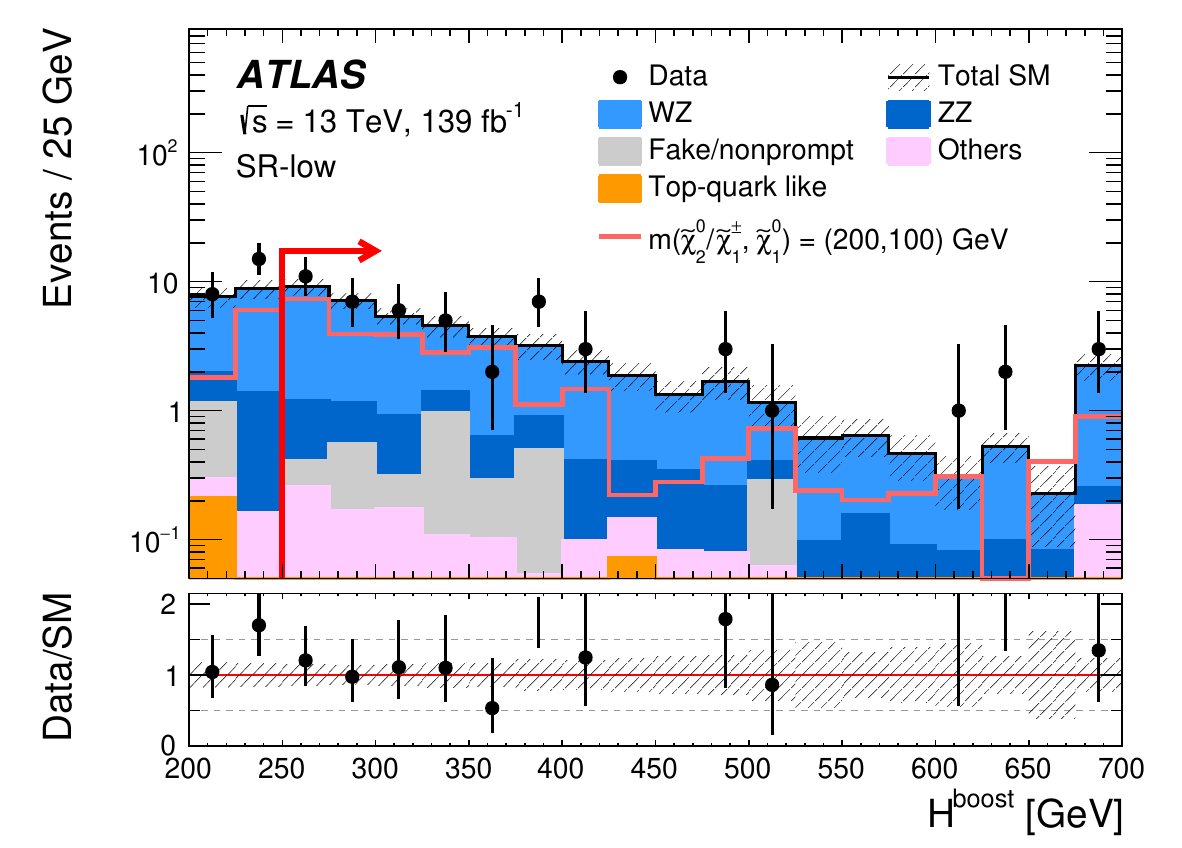}
\label{fig:SRLow:hboost}
}
\subfigure[]{
\includegraphics[width=0.48\textwidth]{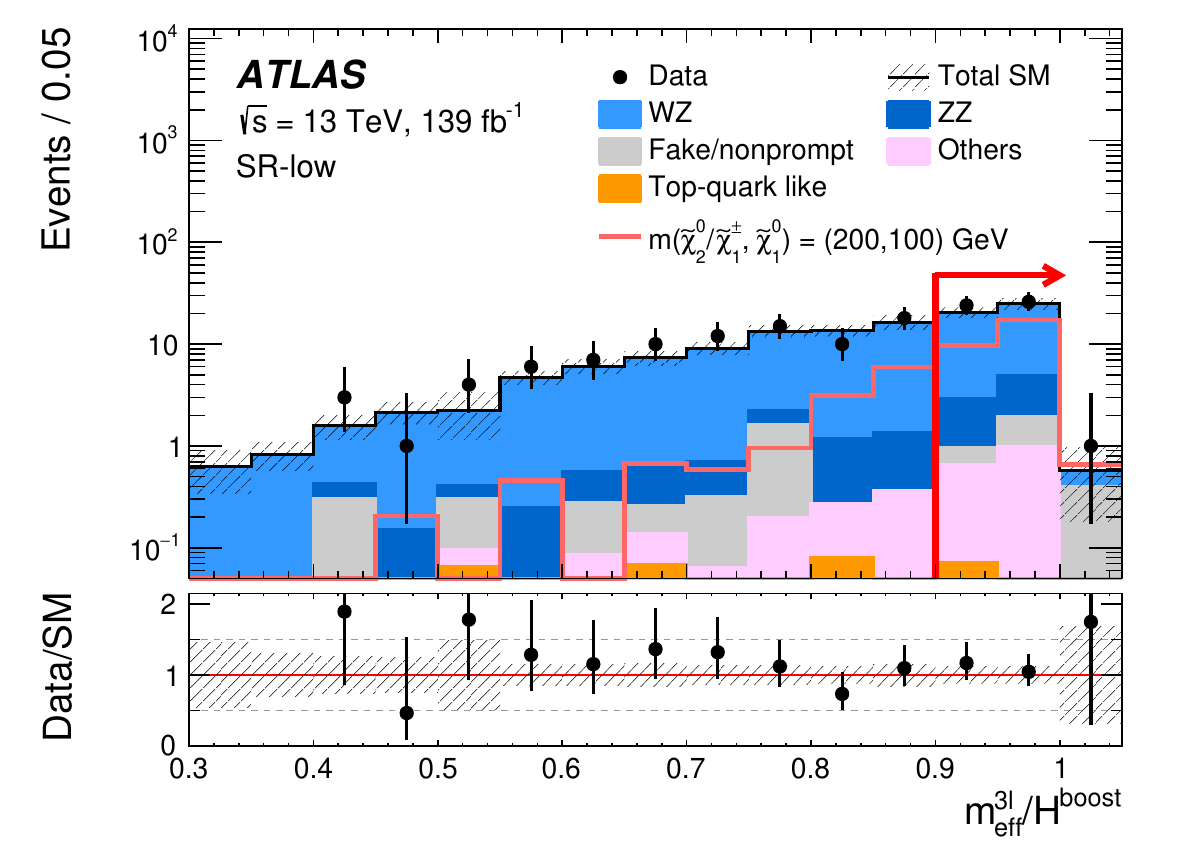}
\label{fig:SRLow:hratio}
}
\subfigure[]{
\includegraphics[width=0.48\textwidth]{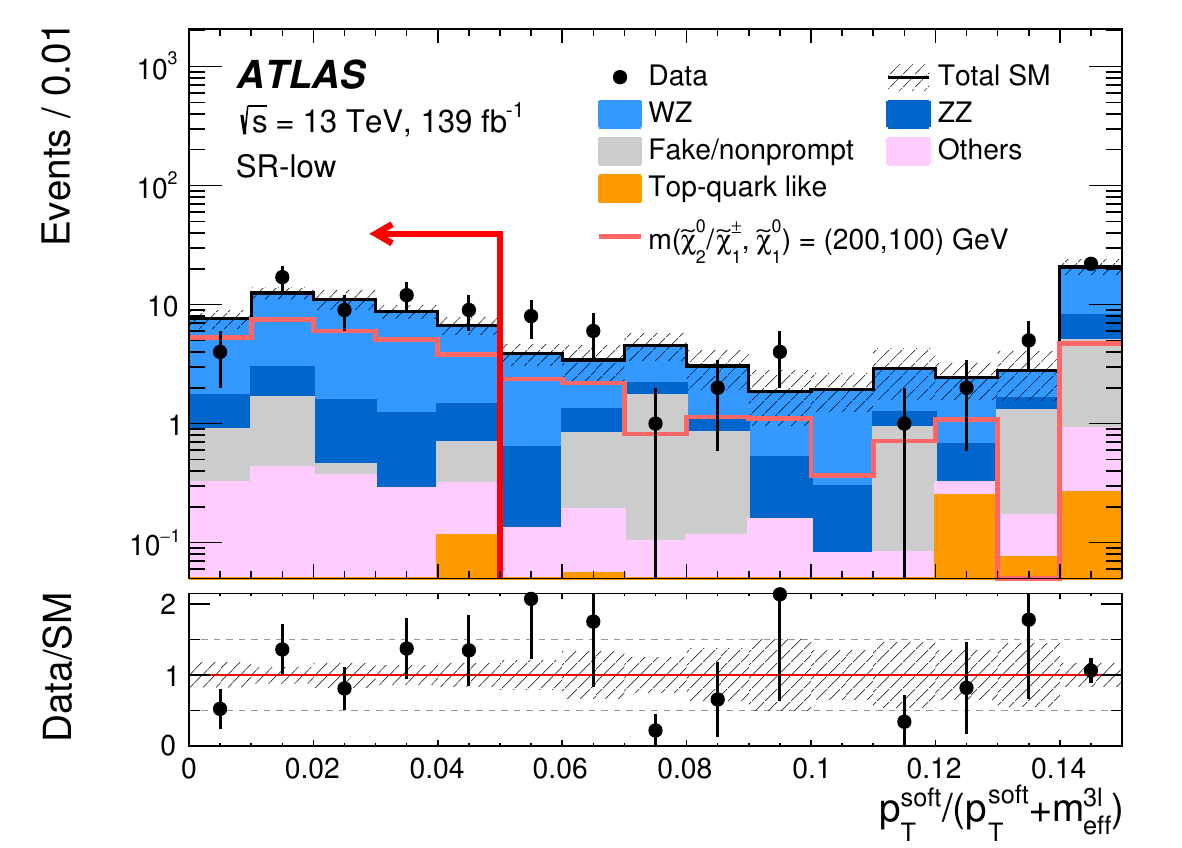}
\label{fig:SRLow:ptratio}
}
 
\caption{Distributions in \SRlow of the data and post-fit background prediction for \subref{fig:SRLow:mt} \mt, \subref{fig:SRLow:hboost} \Hboost, \subref{fig:SRLow:hratio} \inlineHTratio, and \subref{fig:SRLow:ptratio} \inlinepTsoftratio.
\captionsel{\SRlow}
\captionbkgnorm
\captionsig
\captionstyle
}
\label{fig:SRLow}
\end{figure}
 
The HistFitter package~\cite{Baak:2014wma} is used to compute the statistical interpretation based on a log-likelihood method~\cite{Cowan:2010js}. All the systematic uncertainties are treated as Gaussian nuisance parameters in the likelihood.
 
To determine the background prediction, the control regions are used to constrain the $WZ$ normalization assuming no signal events in the CR, referred to as a background-only fit. Normalization factors for the $WZ$ MC simulation are derived from a simultaneous background-only fit of the two orthogonal CRs with all other background processes held constant. The normalization factors are found to be $0.84 \pm 0.07$ for \CRlow and $0.95 \pm 0.05$ for \CRISR.
The two normalization factors are compatible within their uncertainties (described in Section~\ref{sec:systematics}), with small differences due to the modeling of higher-order radiation in the electroweak $WZ$ process.

The observed event yields in the low-mass and ISR regions are compared with the fitted background estimates derived from the log-likelihood fits in Table~\ref{tab:resultsSR} and visualized alongside the validation regions in Figure~\ref{fig:region_yields}.
The data agree well with the background estimates in both signal regions.
Kinematic distributions for these SRs are shown in Figures~\ref{fig:SRLow} and \ref{fig:SRISR}, demonstrating good agreement between the data and the background estimates in the SRs and across the boundaries of the SR selections.

\begin{figure}[h]
\centering
\subfigure[]{
\includegraphics[width=0.48\textwidth]{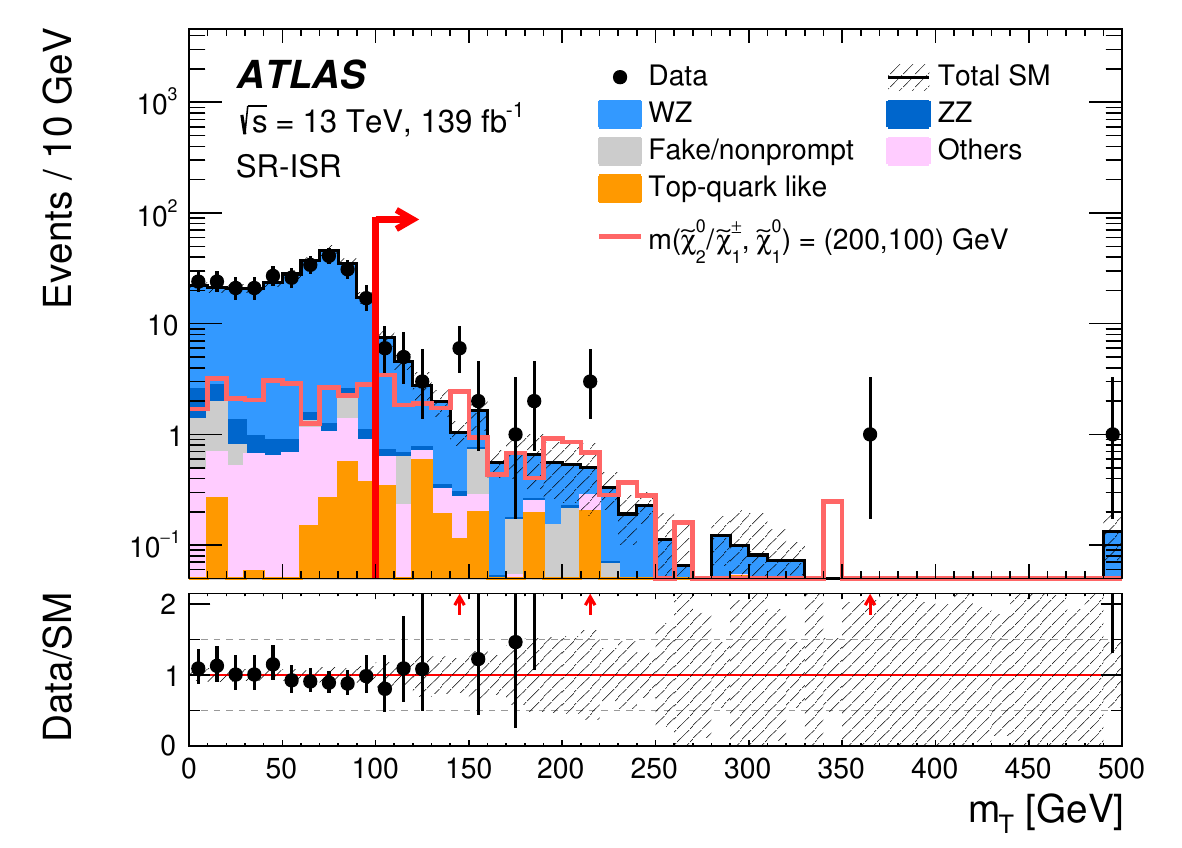}
\label{fig:SRISR:mt}
}
\subfigure[]{
\includegraphics[width=0.48\textwidth]{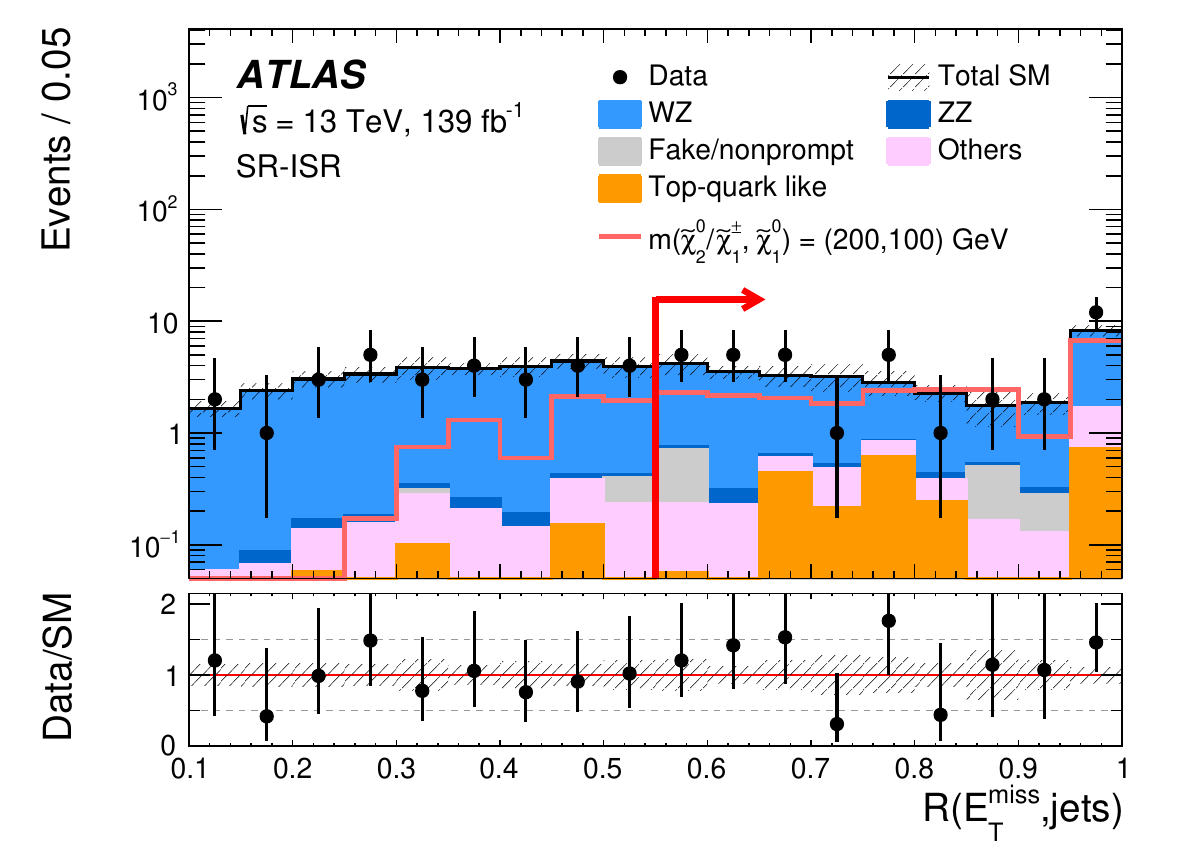}
\label{fig:SRISR:metratio}
}
\subfigure[]{
\includegraphics[width=0.48\textwidth]{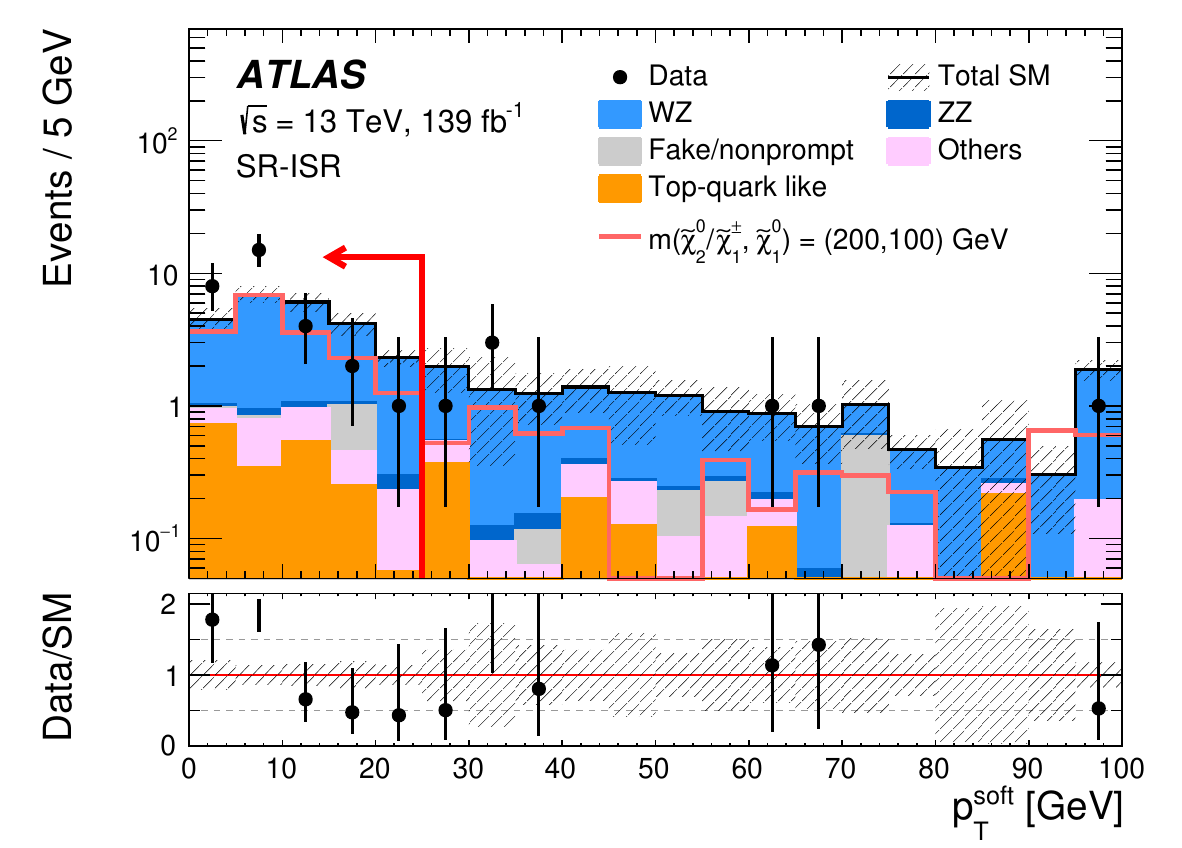}
\label{fig:SRISR:ptsoft}
}
\subfigure[]{
\includegraphics[width=0.48\textwidth]{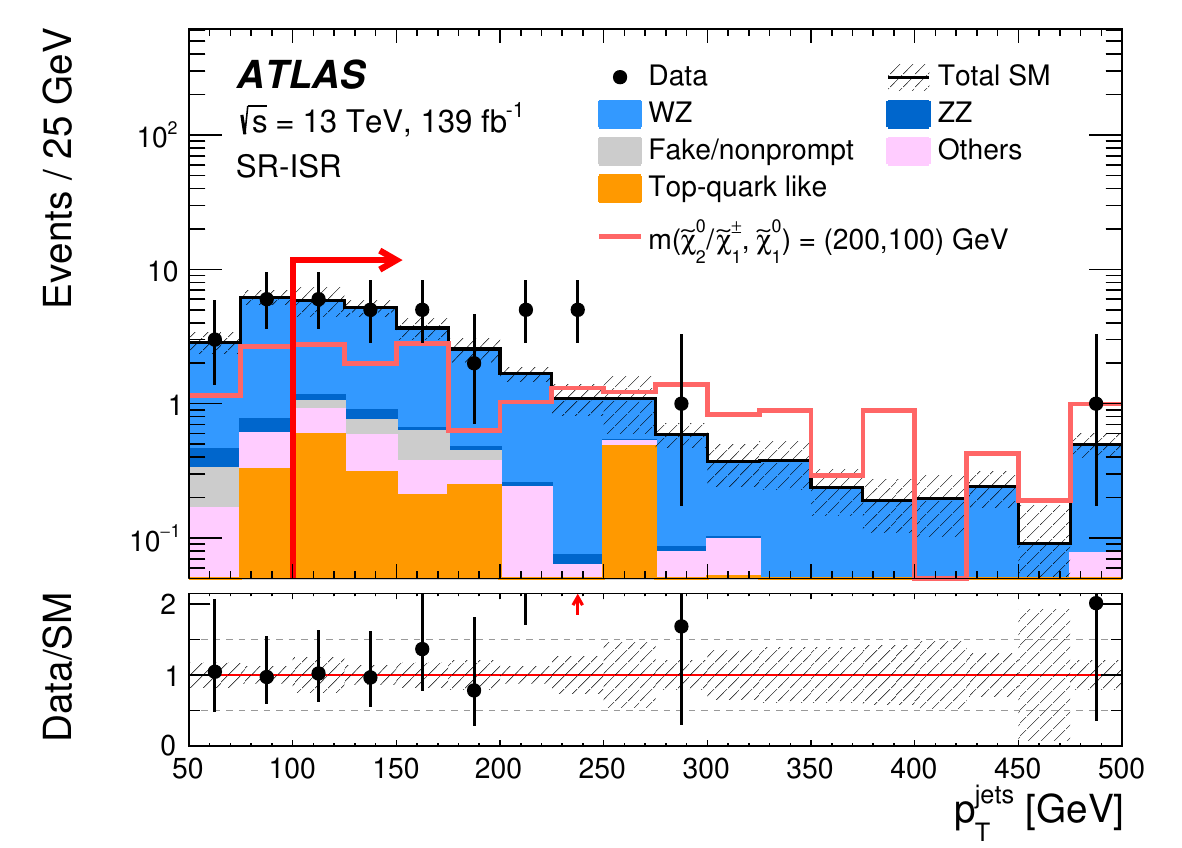}
\label{fig:SRISR:ptjets}
}
 
\caption{Distributions in \SRISR of the data and post-fit background prediction for \subref{fig:SRISR:mt} \mt, \subref{fig:SRISR:metratio} \RMetJets, \subref{fig:SRISR:ptsoft} \pTsoft, and \subref{fig:SRISR:ptjets} \pTjets.
\captionsel{\SRISR}
\captionbkgnorm
\captionsig
\captionstyle
}
\label{fig:SRISR}
\end{figure}
 
As no significant excess is observed, model-independent limits are derived at the 95\% confidence level (C.L.) using the \CLs prescription~\cite{Read:2002hq}. An upper limit on the visible cross section of beyond-the-SM (BSM) processes is derived for each SR. A log-likelihood fit is applied to the number of observed events in the target SR and the associated CR, and a generic BSM process is assumed to contribute to the SR only.
No uncertainties are considered for the signal model except the luminosity uncertainty.
The observed ($S^{95}_\text{obs}$) and expected ($S^{95}_\text{exp}$) limits on the number of BSM events are shown in Table~\ref{table.results.exclxsec.pval.upperlimit.SR}.
Also shown are the observed limits on the visible cross section $\sigma_\text{vis}$, defined as $S^{95}_\text{obs}$ normalized to the integrated luminosity,
which represents the product of the production cross section, acceptance, and selection efficiency of a generic BSM signal.
Limits on $\sigma_\text{vis}$ are set at 0.16~fb in \SRlow and 0.13~fb in \SRISR.
The $p$-value, representing the probability of the SM background alone fluctuating to at least the observed number of events, and the associated significance $Z$ are also shown.
 
\begin{table}
\centering
\caption[Breakdown of upper limits.]{
Summary of the expected background and data yields in \SRlow and \SRISR.
The second and third columns show the data and total expected background with systematic uncertainties.
The fourth column gives the model-independent upper limits at 95\% C.L. on the visible cross section ($\sigma_\text{vis}$).
The fifth and sixth columns give the visible number of observed ($S^{95}_\text{obs}$) and expected ($S^{95}_\text{exp}$) events of a generic beyond-the-SM process,
where uncertainties in $S^{95}_\text{exp}$ reflect the $\pm 1 \sigma$ uncertainties of the background estimates.
The last column shows the discovery $p$-value and Gaussian significance $Z$ assuming no signal.
\label{table.results.exclxsec.pval.upperlimit.SR}}
\setlength{\tabcolsep}{0.0pc}
\begin{tabular*}{\textwidth}{@{\extracolsep{\fill}}lcccccc}
\noalign{\smallskip}\hline\noalign{\smallskip}
{Signal channel}    & $N_\text{obs}$   & $ N_\text{exp}$   & $\sigma_\text{vis}$[fb]  &  $S_\text{obs}^{95}$  & $S_\text{exp}^{95}$  & $p(s=0)$ ($Z$)  \
\\
\noalign{\smallskip}\hline\noalign{\smallskip}
\SRlow    & 51 & 46 $\pm$ 5 &    $0.16$ &  $22.1$ & $ { 19.9 }^{ +7.8 }_{ -3.6 }$& $ 0.27$~$(0.61)$   \\%
\noalign{\smallskip}\hline\noalign{\smallskip}
\SRISR    & 30 & $23.4 \pm 2.1$  & $0.13$ &  $17.8$ & $ { 12.0 }^{ +5.3 }_{ -1.8 }$ & $ 0.11$~$(1.21)$\\%
\noalign{\smallskip}\hline\noalign{\smallskip}
\end{tabular*}
\end{table}

Exclusion limits are derived at 95\% C.L. for models in which pair-produced \chinoonepmninotwo decay exclusively into two \ninoone LSPs, a $W$ boson and a $Z$ boson.
Limits are obtained through a profile log-likelihood ratio test using the \CLs prescription,
following the simultaneous fit to the low-mass and ISR CRs and SRs~\cite{Baak:2014wma}. The signal models are accounted for in this likelihood in both the CRs and SRs.
The low-mass and ISR regions do not affect the nominal fit in the other region due to their orthogonality, but uncertainties that are correlated across regions may be constrained.
Experimental uncertainties are treated as correlated between signal and background events and across low-mass and ISR regions.
The theoretical uncertainty of the signal cross section is accounted for by repeating the limit-setting procedure with the varied signal cross sections and reporting the effect on the observed limit.
 
The expected and observed exclusion contours as a function of the signal \chinoonepm/\ninotwo and LSP \ninoone masses are shown in Figure~\ref{fig:pull_plot_summary_yields_isr}.
Masses can be excluded when the $Z/W$ bosons of the decay are on the mass shell, such that the mass splittings $\Delta m$ are close to or larger than the $Z$ boson mass.
Signal \chinoonepm/\ninotwo events are excluded for masses up to 345~\GeV\ for small \ninoone masses, for which $\Delta m$ is large.
 
\begin{figure}
\centering
\includegraphics[width=0.9\textwidth]{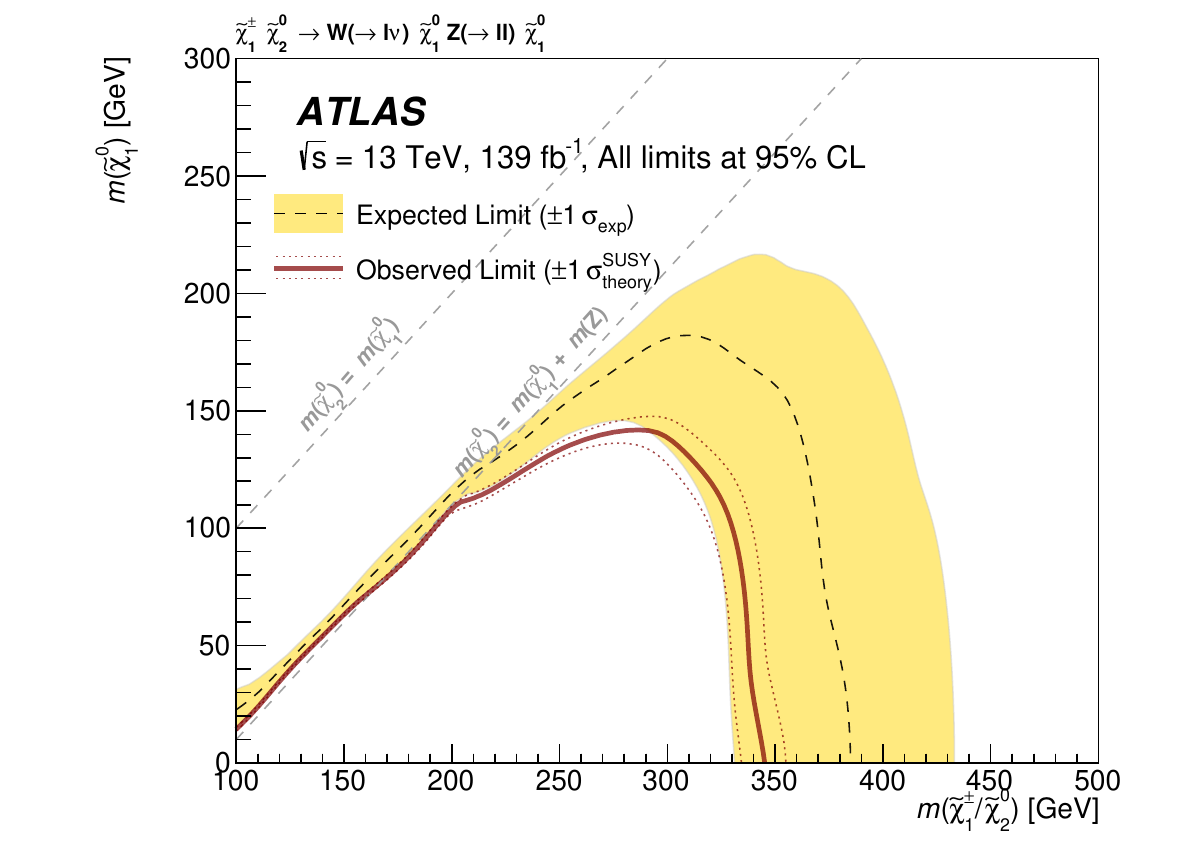}
\caption{
Expected (dashed black) and observed (solid red) exclusion contours for \chinoonepmninotwo production assuming on-shell $W/Z$ decays
as a function of the \chinoonepm/\ninotwo and \ninoone masses, and derived from the combined fit of low-mass and ISR regions.
The yellow band reflects the $\pm 1 \sigma$ uncertainty of the expected limits due to uncertainties in the background prediction and experimental uncertainties affecting the signal.
The dotted red lines correspond to the $\pm 1 \sigma$ cross section uncertainty of the observed limit derived by varying the signal cross section within its uncertainty.
}
\label{fig:pull_plot_summary_yields_isr}
\end{figure}
 
These results extend the exclusion limits in the low-mass and ISR regions beyond those of the RJR analysis from Ref.~\cite{SUSY-2017-03}.
The excesses from the RJR analysis were validated in the $36~\ifb$ of data from the 2015 and 2016 data sets
and found to be reduced with the inclusion of $103~\ifb$ of data from the 2017 and 2018 data sets,
corresponding to local significances of 0.6$\sigma$ in \SRlow and 1.2$\sigma$ in \SRISR.
 

\FloatBarrier
 
\section{Conclusion}
\label{sec:conclusion}
This paper presented a search for pair-produced \chinoonepmninotwo decaying via $W$ and $Z$ bosons into final states with three leptons and missing transverse momentum.
The search targeted electroweakino production for which current limits derived from the recursive jigsaw reconstruction technique and from conventional techniques in the laboratory frame are in tension.
This new search used $139~\ifb$ of proton-proton collisions collected at $\sqs=13~\TeV$ by the ATLAS detector at the LHC between 2015 and 2018.
The data were analyzed with a new emulated recursive jigsaw reconstruction method that uses conventional variables in the laboratory frame to target low-mass electroweakinos and those produced in the presence of initial-state radiation.
A subset of the data collected between 2015 and 2016 was analyzed and excesses were seen for two signal regions of similar construction to those of the recursive jigsaw reconstruction search~\cite{SUSY-2017-03}.
In the full data set the observed event yields were found to be in agreement with Standard Model expectations, with no significant excess seen in either signal region.
The results were interpreted with simplified models of electroweakino pair production, excluding neutralinos and charginos with masses up to 345~\GeV\ at the 95\% confidence level
when the $W$ and $Z$ bosons are on the mass shell.
 
\begin{DIFnomarkup}
\section*{Acknowledgments}
\end{DIFnomarkup}

We thank CERN for the very successful operation of the LHC, as well as the
support staff from our institutions without whom ATLAS could not be
operated efficiently.
 
We acknowledge the support of ANPCyT, Argentina; YerPhI, Armenia; ARC, Australia; BMWFW and FWF, Austria; ANAS, Azerbaijan; SSTC, Belarus; CNPq and FAPESP, Brazil; NSERC, NRC and CFI, Canada; CERN; CONICYT, Chile; CAS, MOST and NSFC, China; COLCIENCIAS, Colombia; MSMT CR, MPO CR and VSC CR, Czech Republic; DNRF and DNSRC, Denmark; IN2P3-CNRS and CEA-DRF/IRFU, France; SRNSFG, Georgia; BMBF, HGF and MPG, Germany; GSRT, Greece; RGC and Hong Kong SAR, China; ISF and Benoziyo Center, Israel; INFN, Italy; MEXT and JSPS, Japan; CNRST, Morocco; NWO, Netherlands; RCN, Norway; MNiSW and NCN, Poland; FCT, Portugal; MNE/IFA, Romania; MES of Russia and NRC KI, Russia Federation; JINR; MESTD, Serbia; MSSR, Slovakia; ARRS and MIZ\v{S}, Slovenia; DST/NRF, South Africa; MINECO, Spain; SRC and Wallenberg Foundation, Sweden; SERI, SNSF and Cantons of Bern and Geneva, Switzerland; MOST, Taiwan; TAEK, Turkey; STFC, United Kingdom; DOE and NSF, United States of America. In addition, individual groups and members have received support from BCKDF, CANARIE, Compute Canada and CRC, Canada; ERC, ERDF, Horizon 2020, Marie Sk{\l}odowska-Curie Actions and COST, European Union; Investissements d'Avenir Labex, Investissements d'Avenir Idex and ANR, France; DFG and AvH Foundation, Germany; Herakleitos, Thales and Aristeia programmes co-financed by EU-ESF and the Greek NSRF, Greece; BSF-NSF and GIF, Israel; CERCA Programme Generalitat de Catalunya and PROMETEO Programme Generalitat Valenciana, Spain; G\"{o}ran Gustafssons Stiftelse, Sweden; The Royal Society and Leverhulme Trust, United Kingdom.
 
The crucial computing support from all WLCG partners is acknowledged gratefully, in particular from CERN, the ATLAS Tier-1 facilities at TRIUMF (Canada), NDGF (Denmark, Norway, Sweden), CC-IN2P3 (France), KIT/GridKA (Germany), INFN-CNAF (Italy), NL-T1 (Netherlands), PIC (Spain), ASGC (Taiwan), RAL (UK) and BNL (USA), the Tier-2 facilities worldwide and large non-WLCG resource providers. Major contributors of computing resources are listed in Ref.~\cite{ATL-GEN-PUB-2016-002}.
 

\clearpage
\printbibliography

\clearpage 
 
\begin{flushleft}
\hypersetup{urlcolor=black}
{\Large The ATLAS Collaboration}

\bigskip

\AtlasOrcid[0000-0002-6665-4934]{G.~Aad}$^\textrm{\scriptsize 102}$,    
\AtlasOrcid[0000-0002-5888-2734]{B.~Abbott}$^\textrm{\scriptsize 129}$,    
\AtlasOrcid{D.C.~Abbott}$^\textrm{\scriptsize 103}$,    
\AtlasOrcid[0000-0002-2788-3822]{A.~Abed~Abud}$^\textrm{\scriptsize 71a,71b}$,    
\AtlasOrcid{K.~Abeling}$^\textrm{\scriptsize 53}$,    
\AtlasOrcid{D.K.~Abhayasinghe}$^\textrm{\scriptsize 94}$,    
\AtlasOrcid[0000-0002-8496-9294]{S.H.~Abidi}$^\textrm{\scriptsize 167}$,    
\AtlasOrcid[0000-0002-8279-9324]{O.S.~AbouZeid}$^\textrm{\scriptsize 40}$,    
\AtlasOrcid{N.L.~Abraham}$^\textrm{\scriptsize 156}$,    
\AtlasOrcid[0000-0001-5329-6640]{H.~Abramowicz}$^\textrm{\scriptsize 161}$
% ...
\bigskip
\\

% ...
$^{40}$Niels Bohr Institute, University of Copenhagen, Copenhagen; Denmark.\\
% ...
$^{53}$II. Physikalisches Institut, Georg-August-Universit\"{a}t G\"ottingen, G\"ottingen; Germany.\\
% ...
$^{71}$$^{(a)}$INFN Sezione di Pavia;$^{(b)}$Dipartimento di Fisica, Universit\`a di Pavia, Pavia; Italy.\\
% ...
$^{94}$Department of Physics, Royal Holloway University of London, Egham; United Kingdom.\\
% ...
$^{102}$CPPM, Aix-Marseille Universit\'e, CNRS/IN2P3, Marseille; France.\\
$^{103}$Department of Physics, University of Massachusetts, Amherst MA; United States of America.\\
% ...
$^{129}$Homer L. Dodge Department of Physics and Astronomy, University of Oklahoma, Norman OK; United States of America.\\
% ...
$^{156}$Department of Physics and Astronomy, University of Sussex, Brighton; United Kingdom.\\
% ...
$^{161}$Raymond and Beverly Sackler School of Physics and Astronomy, Tel Aviv University, Tel Aviv; Israel.\\
% ...
$^{167}$Department of Physics, University of Toronto, Toronto ON; Canada.\\
% ...
$^{a}$ Also at Borough of Manhattan Community College, City University of New York, New York NY; United States of America.\\
$^{b}$ Also at CERN, Geneva; Switzerland.\\
% ...

\end{flushleft}


\end{document}